\documentclass[12pt]{spieman}  
\usepackage{aas_macros}
\usepackage{amsmath,amsfonts,amssymb}
\usepackage{bm}
\usepackage{graphicx}
\usepackage{setspace}
\usepackage{xspace}
\usepackage{tocloft}
\usepackage{longtable}
\usepackage{threeparttable}
\usepackage{booktabs}
\usepackage{caption}
\usepackage{subcaption} 
\usepackage[countmax]{subfloat}
\usepackage{alltt}
\usepackage{tabularx}
\usepackage{rotating,caption}
\usepackage[bottom]{footmisc}
\usepackage{afterpage}
\usepackage{pdflscape}
\usepackage{gensymb}
\graphicspath{{figures/}}
\usepackage{color, colortbl}
\definecolor{Gray}{gray}{0.9}
\usepackage{lineno}

\newcommand{\Edrift}{$\bm{E}_d$\xspace}
\newcommand{\Fe}{$^{55}$Fe\xspace}

\newcommand{\Hubble}{Hubble\xspace} 
\newcommand{\JWST}{JWST\xspace} 
\newcommand{\Euclid}{Euclid\xspace} 
\newcommand{\WFIRST}{WFIRST\xspace} 

\newcommand{\lco}{{$\lambda_\textrm{co}$}\xspace}
\newcommand{\eg}{{\it e.g.}\xspace}

\newcommand{\etc}{{\it etc.}\xspace}
\newcommand{\etal}{{\it et al.}\xspace}
\newcommand{\swir}{$2.5~\mu\rm m$\xspace}
\newcommand{\mwir}{$5~\mu\rm m$\xspace}
\newcommand{\HgCdTeS}{HgCdTe\xspace}
\newcommand{\HgCdTeL}{Hg$_{1-x}$Cd$_x$Te\xspace}

\hypersetup{breaklinks}
\title{Properties and characteristics of the WFIRST H4RG-10 detectors}
\author[a]{Gregory Mosby, Jr.}
\author[a]{Bernard J. Rauscher}
\author[e]{John Auyeung}
\author[b]{Chris Bennett}
\author[d]{E.S. Cheng}
\author[b]{Stephanie Cheung}
\author[b]{Analia Cillis}
\author[c]{David Content}
\author[b]{Dave Cottingham}
\author[b]{Roger Foltz}
\author[c]{John Gygax}
\author[d]{Robert J. Hill}
\author[a]{Jeffrey W. Kruk}
\author[d]{Jon Mah}
\author[d]{Lane Meier}
\author[b]{Chris Merchant}
\author[b]{Laddawan Miko}
\author[e]{Eric C. Piquette}
\author[b]{Augustyn Waczynski}
\author[b]{Yiting Wen}

\affil[a]{NASA Goddard Space Flight Center, Observational Cosmology Laboratory, 8800 Greenbelt Rd, Greenbelt, MD, USA, 20771}
\affil[b]{NASA Goddard Space Flight Center, Detector Systems Branch, 8800 Greenbelt Rd, Greenbelt, MD, USA, 20771}
\affil[c]{NASA Goddard Space Flight Center, Wide Field Infrared Survey Telescope project (Code 448), 8800 Greenbelt Rd, Greenbelt, MD, USA, 20771}
\affil[d]{Conceptual Analytics LLC, 8209 Woburn Abbey Rd, Glenn Dale, MD 20769}
\affil[e]{Teledyne Imaging Sensors, 5212 Verdugo Way, Camarillo, CA, USA 93012}

\cftpagenumbersoff{figure}
\cftpagenumbersoff{table} 
\begin{document} 
\maketitle
\begin{abstract}
The Wide-Field Infrared Survey Telescope (WFIRST) will answer fundamental questions about the evolution of dark energy over time and expand the catalog of known exoplanets into new regions of parameter space. Using a Hubble-sized mirror and 18 newly developed \HgCdTeS 4K x 4K photodiode arrays (H4RG-10), WFIRST will measure the positions and shapes of hundreds of millions of galaxies, the light curves of thousands of supernovae, and the microlensing signals of over a thousand exoplanets toward the bulge of the Galaxy. These measurements require unprecedented sensitivity and characterization of the Wide Field Instrument (WFI), particularly its detectors. The WFIRST project undertook an extensive detector development program to create focal plane arrays that meet these science requirements. These prototype detectors have been characterized and their performance demonstrated in a relevant space-like environment (thermal vacuum, vibration, acoustic, and radiation testing), advancing the H4RG-10's technology readiness level (TRL) to TRL-6. We present the performance characteristics of these TRL-6 demonstration devices.
\end{abstract}


{\noindent \footnotesize\textbf{*}Gregory Mosby, Jr.,  \linkable{gregory.mosby@nasa.gov} }

\begin{spacing}{1}  

\section{Introduction}\label{sec:intro}
The Wide-Field Infrared Survey Telescope (WFIRST) will be the widest field of view (FoV), fastest near-IR (NIR) survey telescope in space\cite{Spergel-2015,Green-2012}. The wide field of view of the Wide Field Instrument (0.281 deg$^2$) is accomplished with a 2.4-m primary mirror and fast optical system that includes a suite of 18 newly designed 4K x 4K ten micron pitched \HgCdTeS photodiode arrays called the H4RG-10. The science objectives of the WFIRST mission require unprecedented sensitivity and precision requirements for the observatory and detectors.

WFIRST's Wide Field Instrument (WFI) will conduct NIR imaging and spectroscopy wide-area surveys (expected to exceed AB~26, with deep fields going several magnitudes fainter) to investigate the Universe's expansion history, probe the growth of structure in the Universe, and complete a census of exoplanets in the Galaxy. Though the WFI will have similar sensitivity and pixel scale as Hubble's WFC3 instrument, the FoV is $\sim$200 times larger\cite{Akeson-2019}. WFIRST will also host a general observer and archive science program. This comprehensive science program will test possible theories for the expansion of the Universe (e.g. dark energy or a cosmological constant or modified gravity), increase our understanding of exoplanet demographics to test theories of planet formation, and provide a platform for general astrophysics observations. The 4 imaging and spectroscopic surveys are a state-of-the-art implementation in designing a mission that will produce one of the richest and complementary data sets in astrophysics aiming to reduce as many systematics possible below statistical uncertainties\cite{Eifler-2020, Eifler-2020b}. 

The High Latitude Spectroscopic Survey (HLSS) will allow the measurement of millions of galaxy redshifts ($1 < z < 2$) to measure cosmic expansion using baryon acoustic oscillation (BAO) measurements of the angular distance, $D_{A}(z)$, and the Hubble parameter, $H(z)$\cite{Dore-2018}. The survey of galaxy redshifts will also provide constraints on the growth of structure in the universe through redshift space distortion (RSD) measurements of the matter clustering amplitude and its time derivative that can be derived from the same data set. A thorough history and description of these observational techniques to probe cosmic acceleration is contained in Ref.~\citenum{Weinberg-2013}.

The High Latitude Imaging Survey (HLIS) will allow measurement of cosmic shear, galaxy-galaxy lensing, and cluster-galaxy lensing\cite{Troxel-2019,Dore-2018}. These measurements probe the growth of structure in the Universe as the measured weak lensing will depend on the distribution of matter between the telescope and the lensed source. These weak lensing measurements will also depend on the distance between the source and the lensing material, providing a measure of cosmic expansion. Thus, the weak lensing measurements will allow cross-checks with BAO and RSD cosmic expansion and structure growth rates results. These measurements will help assess evidence for a cosmological constant, evolving dark energy, or modified gravity.

The Supernova Survey (SNS) will observe thousands of Type Ia supernovae and use their properties as standard candles to measure luminosity distances as a function of redshift\cite{Hounsell-2018}. Luminosity distance depends on key cosmological parameters of the universe including the evolution of dark energy, so these measurements will provide direct measures of the Universe's expansion and the effects of dark energy on the Universe's expansion\cite{Riess-1998,Perlmutter-1997, Sullivan-2011, Suzuki-2012}.

The Exoplanet Microlensing Survey (EMS) will measure exoplanet to star mass ratios and projected separations in units of the Einstein ring radius\cite{Penny-2019}. These measurements will be derived from the transient magnification of source stars and planets as lensing stars in the Galactic Bulge pass between these exoplanetary systems and the observatory. Unlike other methods in exoplanet detection (e.g. transit and radial velocity methods) microlensing will be sensitive to planets from small to large semi-major axes from their stars\cite{Gaudi-2012,Bennett-2002}. From the collection of detected exoplanetary systems, the EMS will allow the precise construction of a census of exoplanets including those beyond the snowline, Mars-mass objects, and free-floating planets\cite{Penny-2019b}. Planet formation theories make distinguishing predictions for the mass function of planetary systems and the frequency of Mars mass and free-floating planets\cite{Chatterjee-2008,Juric-2008,Barclay-2017}. The measurements from the EMS will help evaluate these theories.

The observations needed to accomplish the ambitious tasks of the WFIRST surveys and future guest observer programs depend on reducing instrument systematics and precise measurements from the Wide Field Instrument’s (WFI) detectors. To maximize performance of the detector system we must in general maximize the signal observed and minimize any noise sources. It follows then that for the WFIRST detectors we want devices with high quantum efficiency (QE) and low read noise. The dark current must also be low to minimize its contribution to the total noise that is observed in an exposure; the dark current will generate a shot noise contribution. Similarly, we want devices with little to no persistence as this spurious current from a previous exposure would also contribute a shot noise and is difficult to model and remove. Lastly and more subtlety, we want little to no crosstalk between pixels, or at least a firm grasp on its properties, as the crosstalk between pixels will potentially degrade the instrument point spread function (PSF) jeopardizing, for example, precise shape measurements for the weak lensing measurements of the HLIS. To develop the next generation of NIR detectors to meet these general requirements, the WFIRST project began the H4RG-10 detector development program. 

\subsection{WFIRST IR Detector Technology Advancement Program}\label{sec:intro:development}
To ensure a successful detector fabrication program, the mission phased the development of the H4RG-10 detector arrays. In this Detector Technology Advancement Program (DTAP), a series of experimental devices were built to optimize the design of the WFIRST detectors. These devices included banded array lots, full array lots and flight yield demonstration lots before beginning the flight lots. The banded array lots allowed the project to test multiple pixel design architectures in 1K x 1K sections on one 4K x 4K device. These banded array lots were used to select a pixel design that optimized device characteristics such as the number of hot pixels, total noise, interpixel capacitance, and persistence. The optimization process included variations in implant and contact size, HgCdTe doping levels, surface passivation, cap layer thickness, and epoxy backfill. 

The full array lots were used to assess in 4K x 4K array format two surface passivations (PV2A and PV3) with the selected pixel architecture from the banded array phase. Teledyne considers the details of the passivations to be proprietary. Two flight demonstration lots were made using these recipes. From the set of full array lot devices, the PV3 passivation performed best and was subsequently selected as the flight lot recipe. Following the selection of the flight recipe, we began the growth, hybridization, and packaging of the flight lot detectors. The final phase of the WFIRST DTAP was to demonstrate array performance in relevant environments to advance Technology Readiness Levels\cite{trl_definitions}.

NASA uses Technology Readiness Levels (TRL) to track technology development. A TRL is an integer in the interval [1,9] that approximately describes how mature a technology is and consequently how much risk it carries. TRL-1 is essentially a journal article describing a new technology concept with supporting data. TRL-9 is mature, flight-proven hardware. Within NASA missions, all key technologies are generally required to have met the exit criteria for TRL-6 in order to pass Preliminary Design Review (PDR).

The \WFIRST IR detector development program matured the \HgCdTeS H4RG-10 from approximately TRL-4 to 6. Table~\ref{tab:trl} shows the formal definitions.\cite{trl_definitions}. For visible and near-IR detectors, TRL-4 means that flight-like detectors exist that are close to meeting science requirements. TRL-5 means that some flight-like detectors meet science requirements. TRL-6 means that flight-like detectors have survived environmental testing for the intended application including thermal cycling, vibration, acoustic, and proton irradiation testing.

\begin{table*}[t]
\centering
\caption{NASA Mid-Range TRL Definitions}\label{tab:trl}
\footnotesize
\begin{tabular}{ccccc}
\hline\hline
TRL& Definition& Hardware Description& Software Description& Exit Criteria\\
\hline
4&
\parbox[t]{.75in}{Component and/or breadboard validation in laboratory environment.}& \parbox[t]{1.7in}{A low fidelity system/component breadboard is built and operated to demonstrate basic functionality and critical test environments, and associated performance predictions are defined relative to the final operating environment.}&
\parbox[t]{1.7in}{Key, functionally critical, software components are integrated, and functionally validated, to establish interoperability and begin architecture development. Relevant Environments defined and performance in this environment predicted.}
&\parbox[t]{1.25in}{Documented test performance demonstrating agreement with analytical predictions. Documented definition of relevant environment.}\\
5&
\parbox[t]{.75in}{Component and/or breadboard validation in relevant environment.}& \parbox[t]{1.7in}{A medium fidelity system/component brassboard is built and operated to demonstrate overall performance in a simulated operational environment with realistic support elements that demonstrates overall performance in critical areas. Performance predictions are made for subsequent development phases.}&
\parbox[t]{1.7in}{End-to-end software elements implemented and interfaced with existing systems/simulations conforming to target environment. End-to-end software system, tested in relevant environment, meeting predicted performance. Operational environment performance predicted. Prototype implementations developed.}
&\parbox[t]{1.25in}{Documented test performance demonstrating agreement with analytical predictions. Documented definition of scaling requirements.}\\
6&
\parbox[t]{.75in}{System/sub-system model or prototype demonstration in an operational environment.}&
\parbox[t]{1.7in}{A high fidelity system/component prototype that adequately addresses all critical scaling issues is built and operated in a relevant environment to demonstrate operations under critical environmental conditions.}&
\parbox[t]{1.7in}{Prototype implementations of the software demonstrated on full-scale realistic problems. Partially integrate with existing hardware/software systems. Limited documentation available. Engineering feasibility fully demonstrated.}
&\parbox[t]{1.25in}{Documented test performance demonstrating agreement with analytical predictions.}\\
\hline
\end{tabular}
\end{table*}

In this paper, we review the relevant detector theory to utilize these new devices and compile the typical properties of the WFIRST H4RG-10 arrays. In Section~\ref{sec:theory}, we discuss the theory that underlies the properties that were measured. In Section \ref{det-per}, we summarize focal plane array performance for the qualified full array devices. We also present some preliminary performance properties of the first flight candidate devices.

\section{Theory}\label{sec:theory}

The \WFIRST science program requires excellent control of NIR detector systematics. To do this, one must first understand how H4RG detectors respond to light. Here we provide a short introduction to some of the expected behaviors, both ideal and non-ideal, and the underlying physical mechanisms. We assume that the reader has a basic familiarity with NIR detector arrays. Readers who desire a more thorough introduction may wish to see Rieke's review article.\cite{Rieke:2007fn}

\WFIRST provides the context for our discussion. Our focus is therefore on building a foundation upon which we ultimately hope to limit \WFIRST's systematic calibration residuals to $\ll 1\%$. Our selection of topics is guided by the many helpful questions that we have received from the \WFIRST Science Investigation Teams over the years. For this reason, our focus is mostly on detector artifacts that are the subjects of active research: \eg persistence, burn-in, and count-rate non-linearity. We devote less space to those that are better understood and built into pipelines today, including \eg dark current and classical non-linearity (CNL; although we still provide references for those who are interested in the more established artifacts).

We begin by describing the H4RG architecture. The H4RG is a member of Teledyne Imaging Sensors' ``HxRG'' product line: HxRG $\in$ $\{$H1R, H1RG, H2RG, H4RG$\}$. All HxRGs have essentially the same architecture at pixel level (although specific parameter values differ). We include WFC3 IR's H1R here because it is nearly identical to the others except that its ROIC lacks the built-in guide window feature (the ``G'' in HxRG). We exclude the older HAWAII-1 and 2 arrays in which the light passed through the detector layer's growth substrate (either sapphire or CdZnTe depending on the process) before entering \HgCdTeS. The substrate is removed prior to applying the anti-reflection (AR) coating in all modern HxRGs that are intended for use in space.

We then describe how H4RGs respond to light. Our presentation follows the order of events experienced by an incident photon. We discuss: the AR coating, the absorption of light, charge diffusion, the brighter-fatter effect (BFE), charge capture and release by defect states, charge integration including CNL, read noise in the contact resistance, CNL in the readout integrated circuit (ROIC), and inter-pixel capacitance. We do not include artifacts arising in the control electronics in this article, although we plan to do so in the future. Some of these effects happen in parallel, or at more than one point in the detection process. For example, CNL originates in the photodiodes, in the ROIC, and in the readout electronics. The effects that are mediated by traps in the pn-junctions happen in parallel; including persistence, burn-in, and count rate non-linearity (CRNL). We would be surprised if there were not correlations between these effects. Nevertheless, we have tried as best we can to describe the inter-dependencies, although much remains to be learned.

Some effects are better understood than others. For those that are still poorly understood, persistence for example, we have tried to provide a snapshot of today's understanding with references to the available literature. As \WFIRST matures, we look forward to learning more!

\subsection{The H4RG Architecture}\label{sec:architecture}

\begin{figure}[t]
\centering
\includegraphics[width=3.125in]{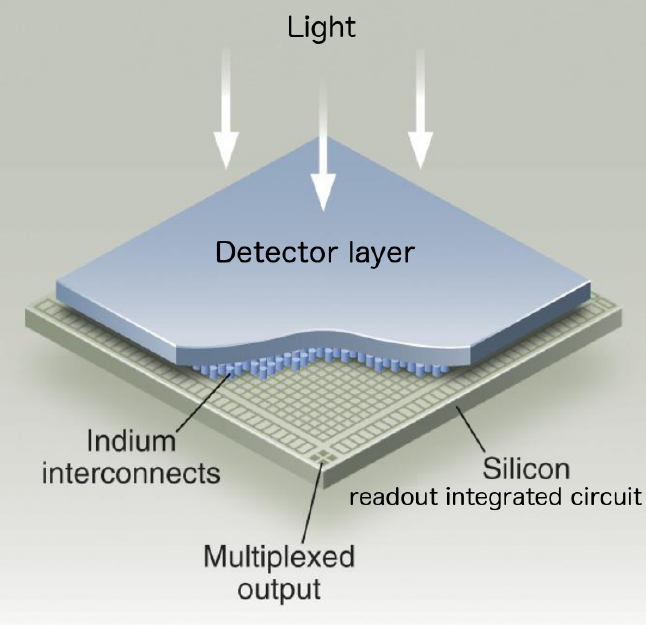}
\caption{The H4RG is a hybrid NIR array detector. The \HgCdTeS detector layer is fabricated separately from the silicon readout integrated circuit. The two are hybridized together using indium interconnects. Although not shown here, for \WFIRST a low viscosity epoxy is wicked between the indium interconnects to increase mechanical strength.}\label{fig:hybrid}
\end{figure}

To understand why a detector behaves the ways that it does, one must understand its architecture. Figure~\ref{fig:hybrid} shows a cutaway view of a generic mercury-cadmium-telluride (\HgCdTeS) NIR array detector.

Here we have used the chemical abbreviation for mercury-cadmium-telluride that one most often sees in the astronomical literature. In the physics literature, one often sees the formula \HgCdTeL, where $x$ specifies the mole fraction of cadmium. In this paper, we use \HgCdTeS, except when discussing the mole fraction of cadmium.

The \HgCdTeS detector layer is ``flip chip'' bonded to the silicon readout integrated circuit (ROIC) using indium solder interconnects. Flip chip hybridization bonding is the name of a microelectronics packaging technique which directly connects an active device to a substrate facedown using solder, thereby eliminating the need for peripheral wirebonds. Although not shown here, for \WFIRST a low viscosity epoxy is wicked between the indium bonds for increased mechanical strength (\JWST and Euclid H2RG arrays use similar backfills). When packaged onto a mechanical mount with electrical connectors and simple fanout circuitry, it becomes a ``sensor chip assembly'' (SCA). 

\begin{figure}[t]
\centering
\includegraphics[width=\textwidth]{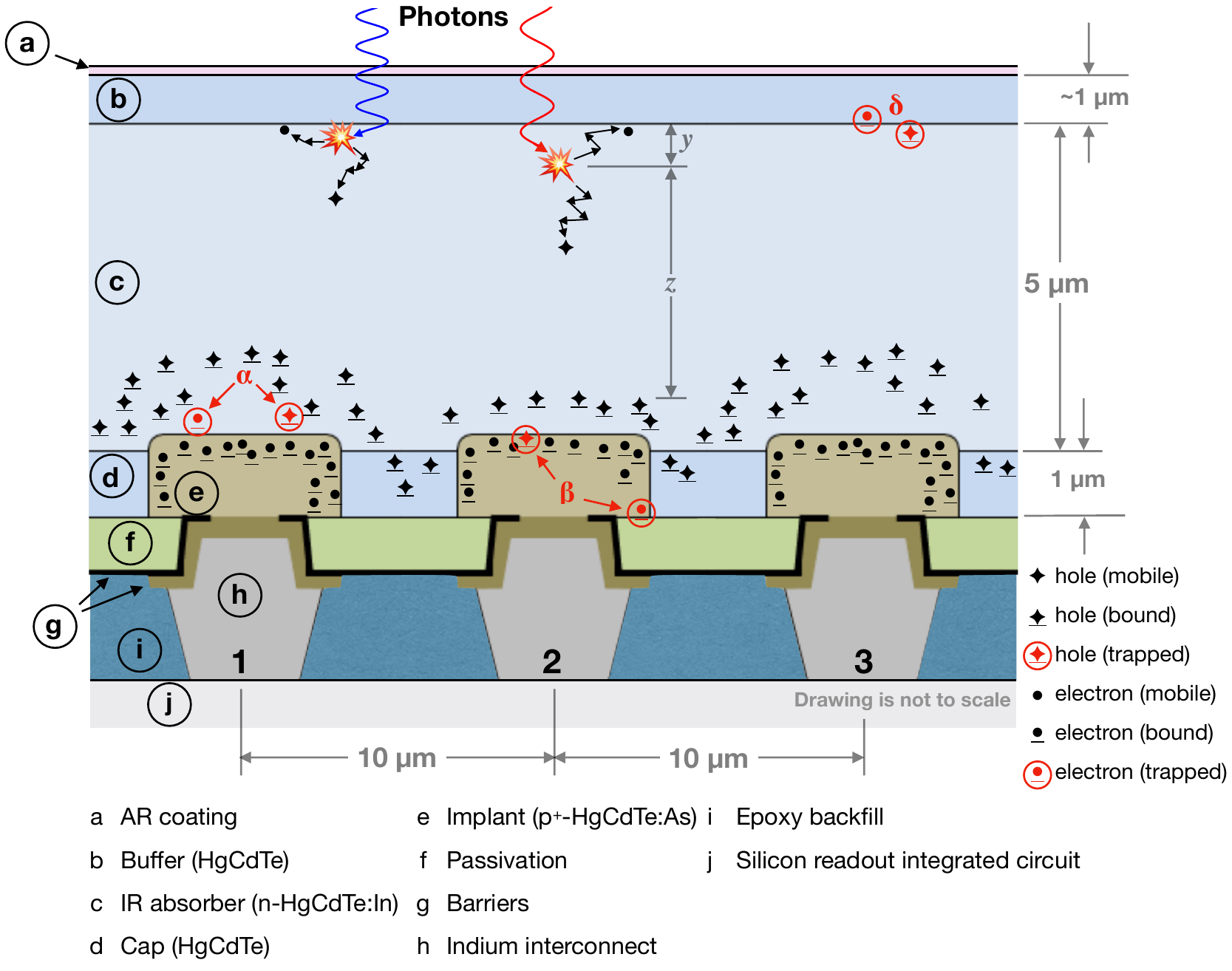}
\caption{\WFIRST's H4RG-10s use a p-on-n \HgCdTeL photodiode architecture. In the ideal case, a photon passes through the AR coating and a transparent buffer before entering the IR absorber. The buffer is needed to transition between the lattice spacing of the (now removed) CdZnTe substrate and the  \HgCdTeL, which depends upon the $x$ parameter. The buffer is transparent because the \HgCdTeS has a wider bandgap than in the absorber. Once in the absorber, the photon excites an electron into conduction creating a mobile electron-hole pair. Following Beer's Law, red light penetrates deeper than blue light. The absorber's bandgap is graded by varying the $x$ parameter. This creates a built in electric field that drives the electron up toward the ``back side'', from where the light entered, and the hole down toward the pn-junctions. Upon reaching depleted \HgCdTeS within about $1.5-2~\mu\rm m$ of the junction (for a freshly reset pixel), the strong electric field there drives the hole into the p-doped implant where it recombines with a bound electron. We return to this figure later to discuss the charge traps and the ``Brigher-Fatter Effect'' that relates to why comparatively full pixel 2 has a smaller collection area than pixels 1 and 3.}\label{fig:photodiode_architecture}
\end{figure}

Figure~\ref{fig:photodiode_architecture} shows a detailed view of the \HgCdTeS detector layer. The detector layer is grown on a CdZnTe substrate using molecular beam epitaxy (MBE). CdZnTe is uses because it provides a good match to the lattice-spacing of \HgCdTeS. This differs from the liquid phase epitaxy (LPE) that was used to grow \HgCdTeS on sapphire substrates for some of the older Rockwell arrays (including those used by Hubble's NICMOS instrument).\cite{Poksheva1993} As will be described later, one advantage that MBE enjoys over LPE is that the bandgap can be tuned during layer growth. The substrate does not appear in Figure~\ref{fig:photodiode_architecture} because it is removed prior to applying the AR coating.

As we discuss the physics of these devices, we will occasionally need to reference the \HgCdTeS's material properties. Table~\ref{tab:hgcdte-params} shows the recommended values. For comparison (and because the information is not published elsewhere), we provide the corresponding values for the James Webb Space Telescope (JWST) Near Infrared Spectrograph (NIRSpec). For more information on \JWST detectors, the interested reader is referred to Ref.~\citenum{Rauscher_2014}.

\begin{table*}[t]
\centering
\caption{\HgCdTeL Parameters}\label{tab:hgcdte-params}
\small
\begin{tabular}{clccl}
\hline\hline
&&\multicolumn{2}{c}{Value}\\
\cline{3-4}
Parameter&Description&WFIRST&JWST$^a$&Comment\\
\hline
$\lambda_{\rm co}$&Cutoff wavelength$^b$&$2.5~\rm \mu m$&$5.3~\rm \mu m$&Science requirement\\
$T$&Operating temperature&$95$&$42.8$&\parbox[t]{1.7in}{Needed to meet dark current requirement}\\
$x$&Cadmium mole fraction&$0.445$&$0.3$&\parbox[t]{1.7in}{Teledyne design value}\\
$\xi$&Pixel pitch&$10~\mu\rm m$&$18~\mu\rm m$&$\dots$\\
$Z$&Absorber thickness&$5~\rm \mu m$&$6~\rm \mu m$&$\dots$\\
$\left|\bm{E}_d\right|$&Drift field strength&$20\ \rm V\ cm^{-1}$&$18\ \rm V\ cm^{-1}$&$\dots$\\
$\epsilon$&Dielectric constant&$14.63$&$16.4$&\\
$L_h$&Hole diffusion length&$20\ \rm \mu m$&$27\ \rm \mu m$&Teledyne measurements\\
$m_h^*$&Hole effective mass&$0.55 m_0$&$0.55 m_0$&Refs.~\citenum{Rogalski_2005, Kinch_2007}\\
$\mu_h$& Hole mobility& $\rm 80~cm^2 V^{-1} s^{-1}$& ---& \parbox[t]{1.7in}{See note $c$}\\
$\rm V_{\rm rst}$&Reset bias&$1\ \rm V$&$0.25\ \rm V$&\parbox[t]{1.7in}{$\rm Vreset-Dsub$, see the Teledyne H4RG user manual}\\
$\sigma_d$&\parbox[t]{1.5in}{Charge diffusion kernel width}&$2.04\ \rm \mu m$&$1.58\ \rm \mu m$&\parbox[t]{1.7in}{Computed using Eq.~\ref{eq:stdev} for very blue in-band light.}\\
\hline
\multicolumn{5}{l}{\parbox[t]{6.3in}{$^a$The values shown are for the Near Infrared Spectrograph (NIRSpec). Many of these values have not previously been published.}}\\
\multicolumn{5}{l}{\parbox[t]{6.3in}{$^b$These are the standard cutoff wavelengths that Teledyne advertises. The physical cutoff of the \HgCdTeS is usually slightly longer to ensure meeting QE requirements at all wavelengths.}}\\
\multicolumn{5}{l}{\parbox[t]{6.3in}{$^c$This is a very rough approximation that follows Refs.~\citenum{Rogalski_2005,Rosbeck_1982} assuming $\mu_h=0.01\times \mu_e$. We have no recommendation for \JWST at this time.}}\\
\end{tabular}
\end{table*}

With MBE-grown \HgCdTeL detector layers, it is possible to engineer the bandgap by varying the $x$ parameter during growth. The standard empirical relation is that of Hansen, Schmit, and Casselman,\cite{Hansen_1982}
\begin{equation}
E_g\left(x,T\right) = -0.302 + 1.93x + 5.35\left(10^{-4}\right)T\left(1-2x\right)\\ -0.810x^2 + 0.832x^3,\label{eq:hansen}
\end{equation}
where $E_g$ is the bandgap energy in eV and $T$ is temperature in degrees Kelvin. The bandgap energy is related to the cutoff wavelength, \lco, by
\begin{equation}
E_g = \frac{h c}{q\lambda_\textrm{co}},
\end{equation}
where $q$ is the elementary charge. For \WFIRST's $\lambda_{\rm co}\approx 2.5~\mu\rm m$ detectors, $x=0.445$, which corresponds to a slightly longer cutoff wavelength for the \HgCdTeS material itself.

\begin{figure}[t]
\centering
\includegraphics[width=5in]{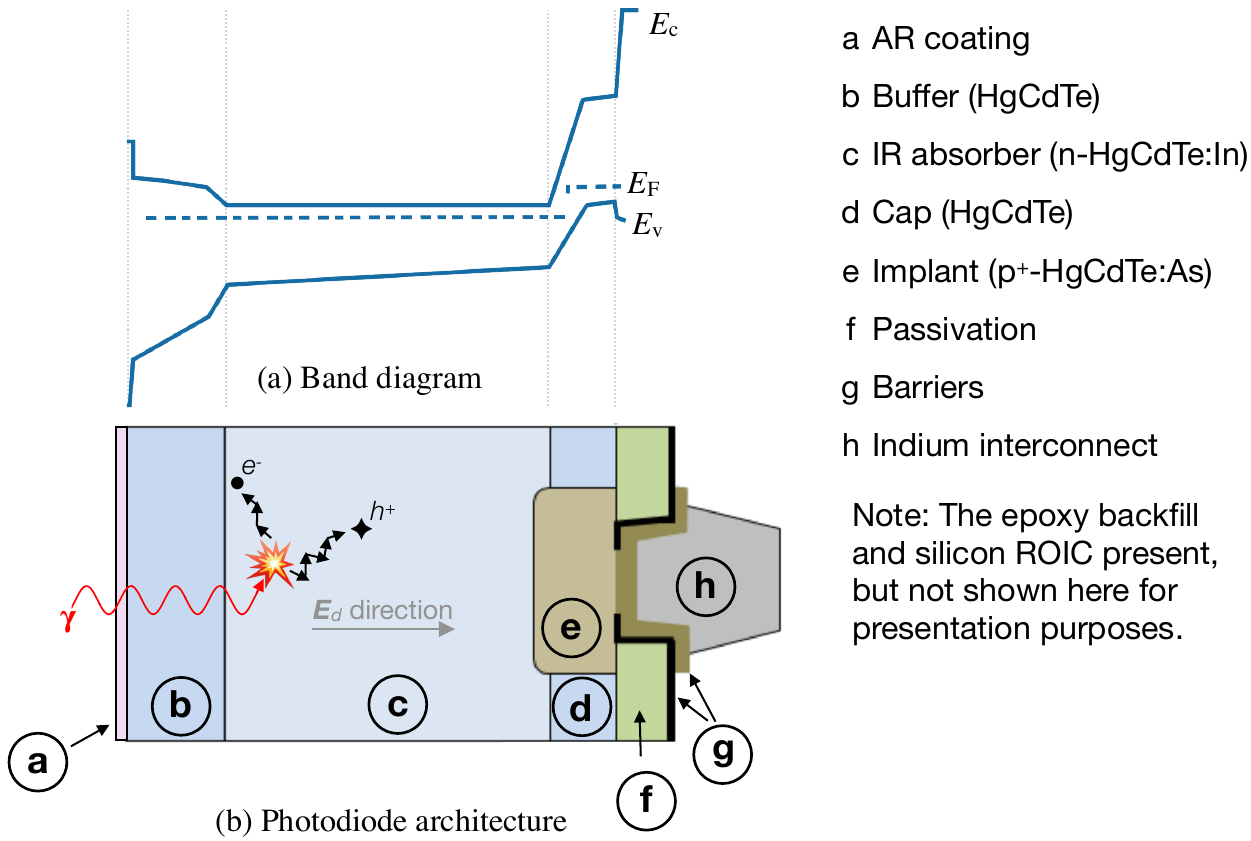}
\caption{(a) MBE enables grading the bandgap by varying the \HgCdTeL's $x$ parameter during layer growth. The bandgap is graded everywhere, thereby creating a grown-in electric field at all places within the \HgCdTeS. Within the IR absorber, this is the drift field of amplitude $\left|\bm{E}_d\right|$. As discussed in the text, the drift field drives photo-generated holes in toward the pn-junction, where they are efficiently collected. For reference, panel (b) shows the photodiode architecture using the same labeling as in Figure~\ref{fig:photodiode_architecture}.}\label{fig:band_diagram}
\end{figure}

The absorber's bandgap is graded by varying the \HgCdTeL's $x$ parameter. Figure~\ref{fig:band_diagram} shows the band diagram. The graded bandgap creates a built in drift field, \Edrift, that drives holes down toward the pn-junctions (here boldface italics denote vectors). For \WFIRST, its amplitude is $\left| \bm{E}_d\right| \approx20~\textrm{V}~\textrm{cm}^{-1}$, which is quite weak compared to the $> 100\times$ stronger effective $\bm{E}$ field that exists in the depleted \HgCdTeS near the pn-junctions immediately after reset. The drift field improves QE and count rate dependent non-linearity (CRNL) by driving holes away from any backside traps that might exist. It reduces lateral charge spreading by reducing the time available for mobile holes to diffuse in the absorber.

During astronomical exposures, including even of very bright objects, there is rarely more than one mobile charge pair in a pixel at any time. Once created, the electrons and holes diffuse in opposite directions under the influence of $\bm{E}_d$. The holes are driven deep into the \HgCdTeS, where nearly all reach depleted \HgCdTeS and are eventually integrated. The electrons are driven back up, toward the entering photons, where they quickly equilibrate with the detector substrate bias voltage.

In the next few sections, we follow the light as it works its way through a \WFIRST detector. The sequence flows from top to bottom of Figure~\ref{fig:photodiode_architecture}.                          

\subsection{Anti-Reflection Coating}\label{sec:theory:ar}

The AR coating is the last optical surface in the WFI. It functions analogously to an interference filter, ensuring high throughput everywhere in the required wavelength range.

\begin{figure}[t]
    \centering
    \begin{subfigure}[t]{.48\textwidth}
        \centering
        \includegraphics[width=\textwidth]{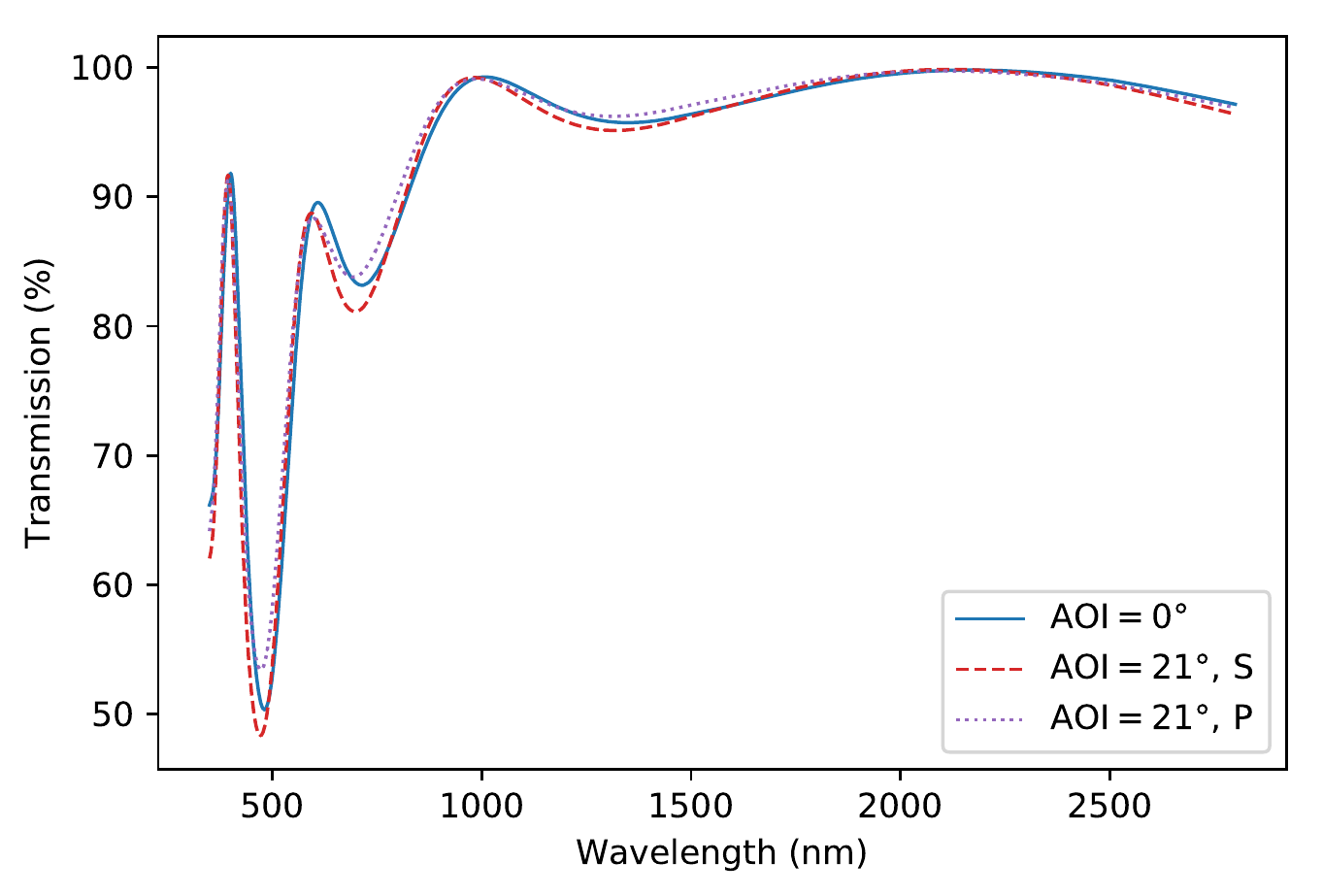} 
        \caption{Differing angle of incidence}
    \end{subfigure}
    \begin{subfigure}[t]{0.48\textwidth}
        \centering
        \includegraphics[width=\textwidth]{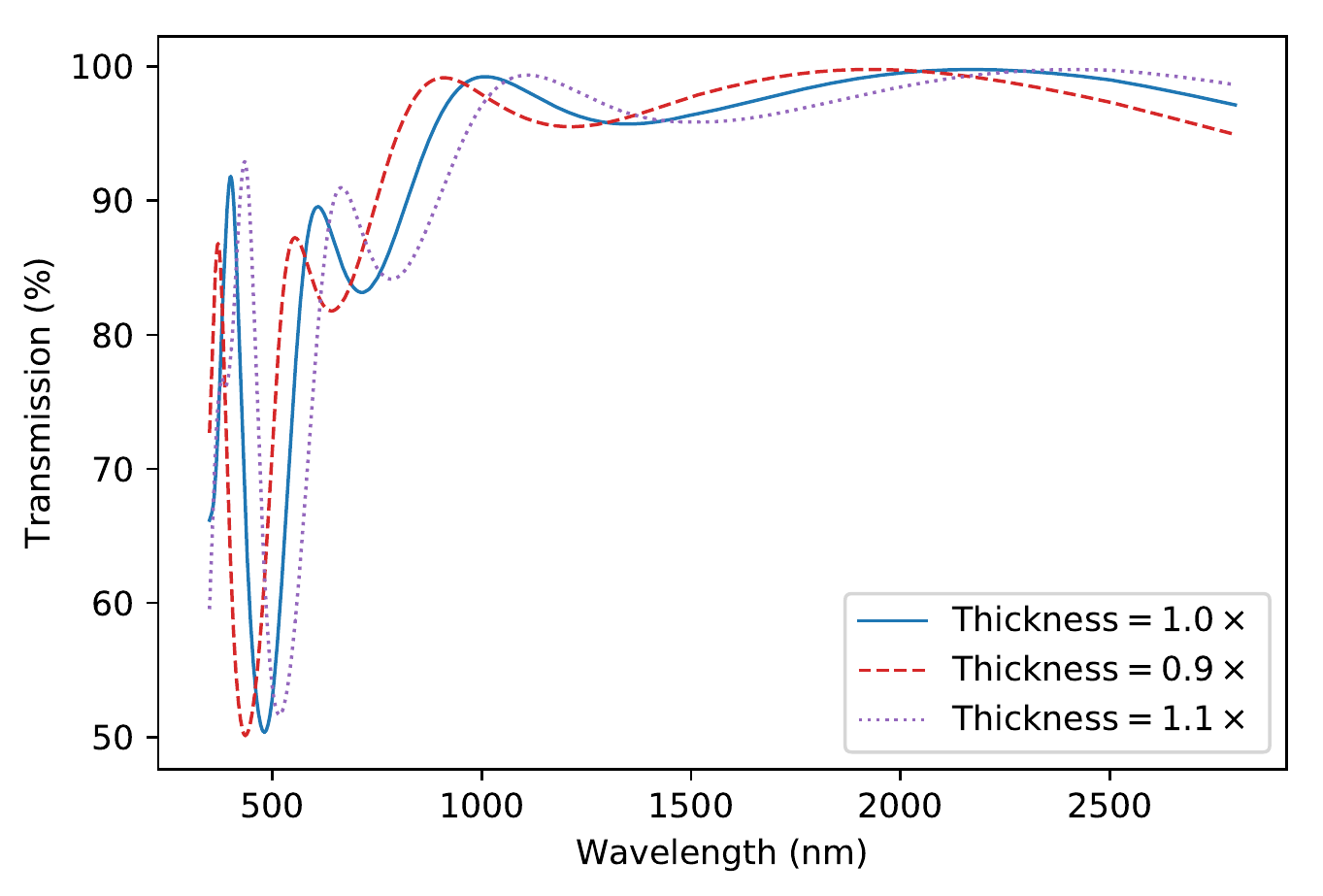} 
        \caption{Differing coating thickness}
    \end{subfigure}
    \caption{This figure shows a), the \WFIRST AR coating model for normal incidence, and also for angle of incidence, $\rm AOI=21\degree$. The ``S'' and ``P'' designations refer to polarization. S-polarized is transverse electric, with the electric field perpendicular to the plane of incidence and P-polarized is transverse magnetic with the electric field in the plane of incidence. The locations of the wiggles do not change greatly, but the coating transmits P-polarized light somewhat better. b) In real devices, both from detector to detector and within an SCA, there will be some variation in the AR coating's thickness. The effects include shifting the positions of the peaks and valleys and changes in the transmission, particularly at blue wavelengths. The sense is that thinner coatings have the peaks and valleys shifted to shorter wavelengths. }\label{fig:ar_coating}
\end{figure}

Because it is based on interference, the coating's transmission has bumps and wiggles (Figure~\ref{fig:ar_coating}). Some of these are large enough to matter for astrophysics. Although the AR coating's design is proprietary, the transmission model results are not. Figure~\ref{fig:ar_coating} shows the coating model for a few incidence angles and hypothetical 10\% deviations from the design thickness. The underlying data, which include more angles of incidence than are plotted here, are available from the \WFIRST web pages.

\subsection{Absorption of Light}\label{sec:theory:absorption}

The transmitted light passes through the transparent \HgCdTeS buffer and into the IR absorber, where nearly all is absorbed. The buffer layer is needed to transition from the lattice spacing of the CdZnTe growth substrate to that of the photosensitive \HgCdTeS. It is transparent because its bandgap exceeds the energy of in-band photons. The CdZnTe substrate does not appear in Figure~\ref{fig:photodiode_architecture} because it is removed prior to applying the AR coating.

The Beer-Lambert law governs extinction within the \HgCdTeS,
\begin{equation}
    I\left(y\right) = I_0 e^{-\alpha y},\label{eq:beer-lambert}
\end{equation}
where $\alpha$ is the extinction coefficient and $y$ (see Figure~\ref{fig:photodiode_architecture}) is depth in the absorber.

Several studies have focused on the extinction of \HgCdTeS near the cutoff wavelength.\cite{Finkman_1979,Finkman_1985} However, most of the time in astronomy we are more interested in the in-band light. Chu~\etal\cite{Chu_1994} give the following empirical relations that include both sub-bandgap and in-band light. In Equations~\ref{eq:chu-7}-\ref{eq:chu-6}, $\alpha$ is measured in cm$^{-1}$ and energy is measured in eV.

For in-band light, $E\geq E_g$, Chu~\etal's relation is,
\begin{equation}
    \alpha = \alpha_g \exp \sqrt{\beta\left(E-E_g\right)}, \textrm{where}\label{eq:chu-7}
\end{equation}
\begin{equation}
    \beta\left(T,x\right) = -1+0.083T+\left(21-0.13T\right)x, \textrm{and}\label{eq:chu-10}
\end{equation}
$\alpha_g$ is given by Eq.~\ref{eq:chu-5}.

For out-of-band light beyond the cutoff, $E<E_g$, Chu~\etal give,
\begin{equation}
    \alpha = \alpha_0\exp \frac{\delta\left(E-E_0\right)}{k T},\label{eq:chu-1}
\end{equation}
\begin{equation}
    \alpha_0 = \exp \left(-18.5+45.68x\right),\label{eq:chu-2}
\end{equation}
\begin{equation}
    E_0 = -0.355+1.77 x,\label{eq:chu-3}
\end{equation}
\begin{equation}
    \frac{\delta}{k T} = \frac{\ln\alpha_g-\ln\alpha_0}{E_g-E_0},\label{eq:chu-4}
\end{equation}
\begin{equation}
    \alpha_g = -65+1.88T+\left(8694-10.31T\right)x, \textrm{and}\label{eq:chu-5}
\end{equation}
\begin{multline}
    E_g\left(x,T\right) = -0.295+1.87x-0.28x^2\\ +\left(6-14x+3x^2\right)\left(10^{-4}\right)T+0.35x^4.\label{eq:chu-6}
\end{multline}
In Equation~\ref{eq:chu-5}, $\alpha_g$ is the absorption coefficient at the bandgap energy. Chu~\etal's empirical relation for $E_g$ differs from Hansen's relation that we otherwise recommend (Eq.~\ref{eq:hansen}). Except when using Chu~\etal's formulae for consistency, we recommend using Equation~\ref{eq:hansen} as it is is more widely accepted.

\begin{figure}[t]
\centering
\includegraphics[width=3.125in]{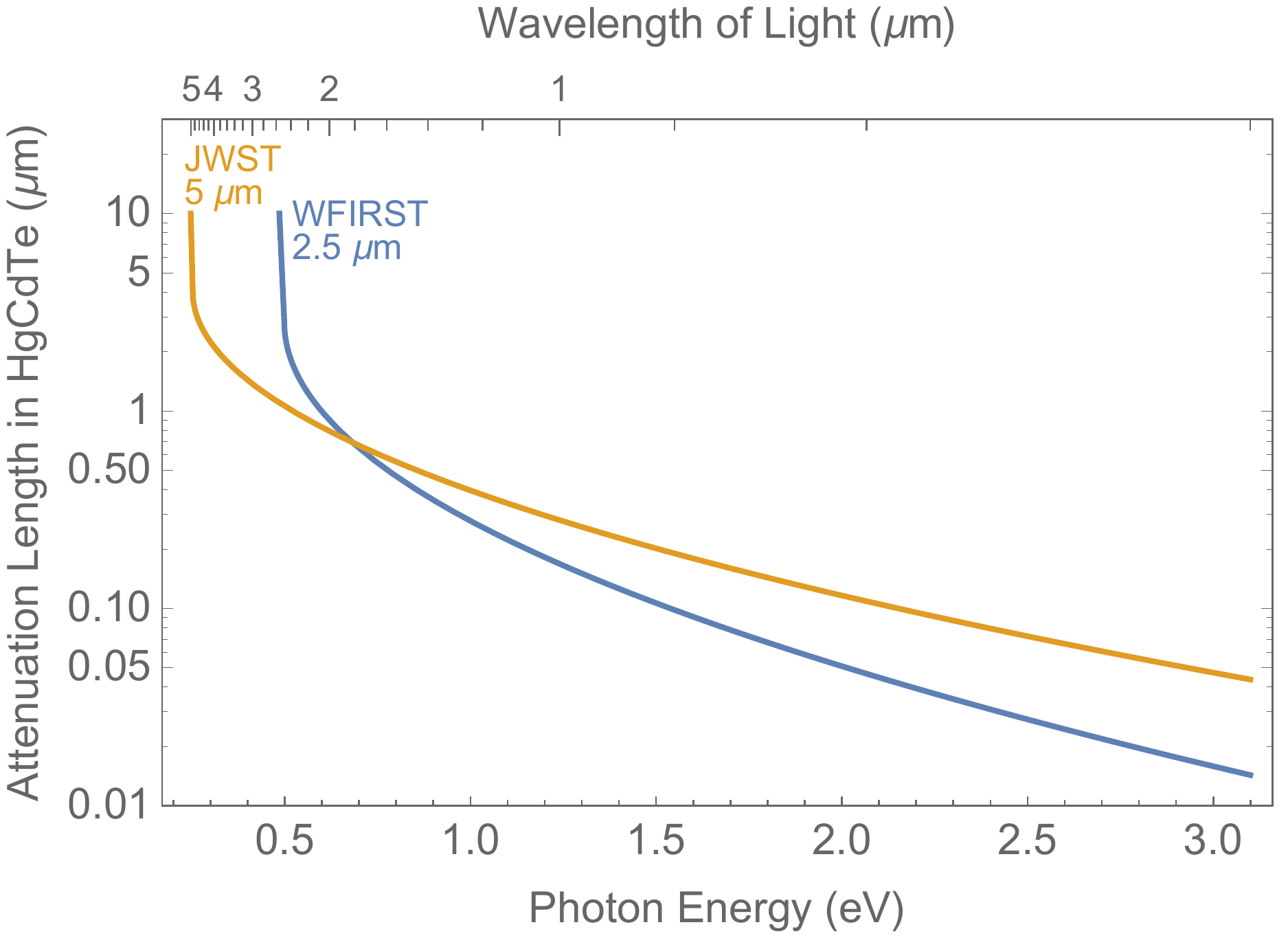}
\caption{\HgCdTeS is extremely opaque to wavelenths shorter than the cutoff wavelength. The Beer-Lambert law governs the extinction as a function of depth. After passing through the AR coating and transparent \HgCdTeS buffer, most light is absorbed within about 1~$\mu$m of the absorber's surface. The buffer layer is transparent because the bandgap there exceeds the energy of in-band photons. Near the cutoff wavelength, the \HgCdTeS rapidly becomes transparent. }\label{fig:attenuation_length}
\end{figure}

The attenuation length is the reciprocal of the absorption coefficient. Figure~\ref{fig:attenuation_length} plots the attenuation length as a function of photon energy and wavelength. Except for wavelengths near the cutoff, most light is absorbed within the first $\approx 1~\mu\rm m$ after entering the absorber.

For \WFIRST, the combination of: (1) \HgCdTeL with $x=0.445$ , (2) a $5~\mu$m thick absorber, and (3)  all planned science being done at wavelengths $\lambda<2~\mu$m means that fringing\cite{Mott2008} is extremely unlikely with a correctly fabricated \WFIRST H4RG.

\subsection{Charge Diffusion}\label{sec:theory:spreading}

After the photon is absorbed, the newly created hole must usually diffuse though several microns of absorber before reaching depleted \HgCdTeS. Fortunately, because of the built-in drift field, charge diffusion degrades image sharpness only slightly in modern HxRG arrays. We expect the WFI's optical PSF and IPC to be more important sources of image blurring in the detector.

Even for cosmic rays, the diffusive spreading will be much less than one would expect if the drift field were not there (for example, the old NICMOS detectors did not have a drift field). This has important implications for space missions because it reduces the number of pixels that are corrupted by cosmic rays.\cite{Pickel2002} Although cosmic rays are not our focus in this paper, we may return to them in future publications.

Several authors have studied charge diffusion in closely-space photodiode arrays.\cite{Kirkpatrick1979,Holloway1986,Pickel2002} Unfortunately, none of these are directly applicable to \WFIRST because they neglect the H4RG's drift field, \Edrift. The drift field drives mobile holes toward the pn-junctions, thereby increasing QE and reducing charge diffusion compared to what it would otherwise be. The drift field improves QE by keeping mobile holes away from backside surface traps. It reduces charge diffusion by quickly sweeping mobile holes out from the absorber layer. \Edrift exists in all modern HxRG detectors; including those used by the \Hubble Wide Field Camera 3 (WFC3), \JWST, \Euclid, and \WFIRST.

Here we show that charge diffusion in modern HxRGs can be approximated using a Gaussian point spread function (PSF; symbol $\Psi$ in equations). By comparing \Fe X-ray crosstalk and electronic IPC measurements, we did a quick check of the Gaussian PSF model for one H4RG. After accounting for IPC, the model correctly predicted  $\approx 0.4\%$ crosstalk to each of the four nearest neighbors. It also significantly under-predicted the measured $0.04\%$ crosstalk to each of the next nearest-neighbors. However, we believe that other crosstalk mechanisms are likely in the available data at the $<0.1\%$ level. We look forward to doing more comprehensive testing as more and better data become available.

Our starting point is the observation that for \WFIRST, the absorber crossing time is only of order,
\begin{equation}
\tau_h = \frac{Z}{\mu_h \left|\bm{E}_d\right|}\approx 0.3~\mu\rm s.\label{eq:crossing_time}
\end{equation}
With this short crossing time, there will seldom be more than one mobile hole in a pixel at any time. A random walk with drift is therefore a more intuitive model than drift-diffusion which posits that mobile holes interact with one another while crossing the absorber.

\subsubsection{Charge Diffusion as a Random Walk}\label{sec:theory:spreading:theory}

\begin{figure}[t]
\begin{center}
\includegraphics[width=3.125in]{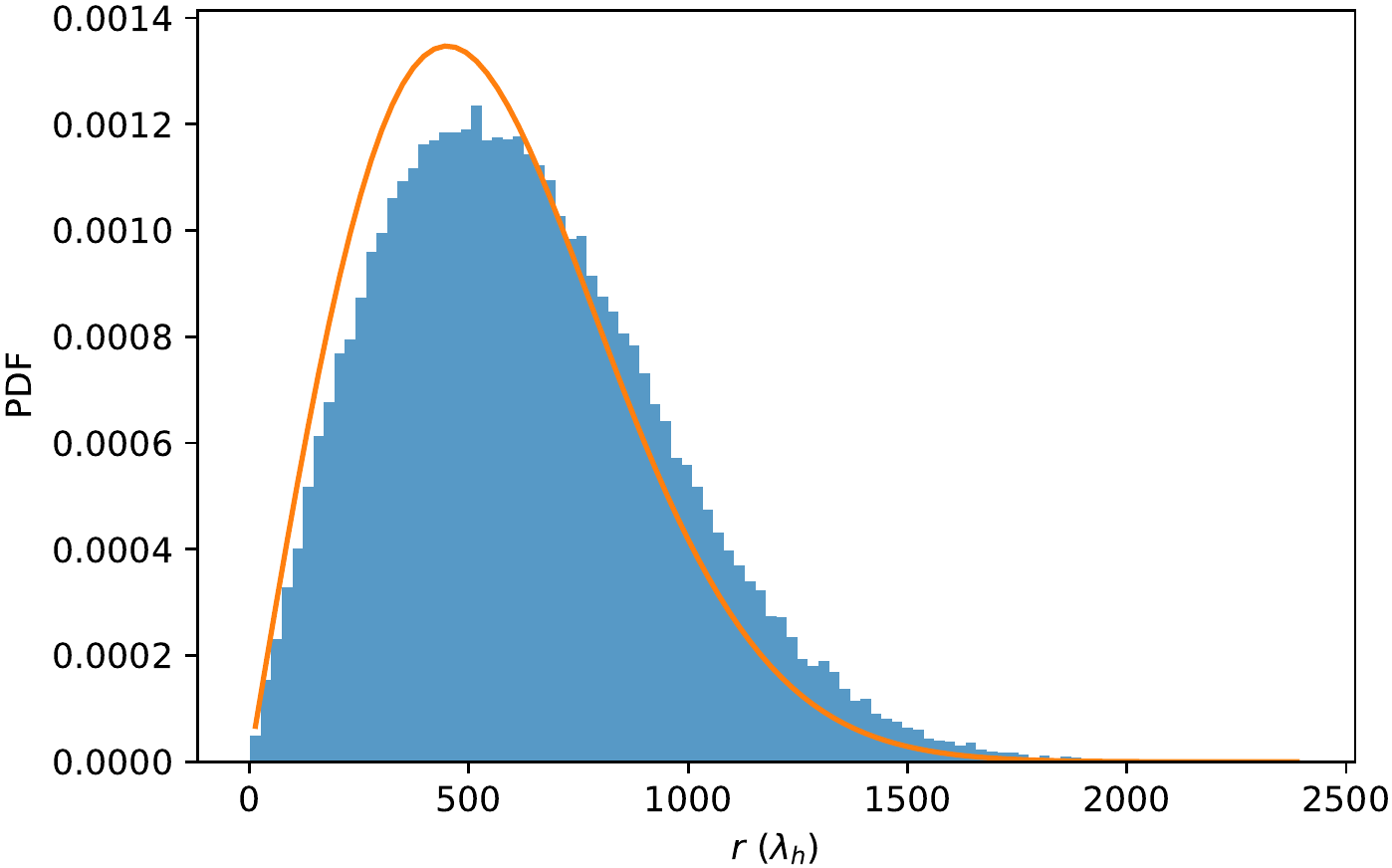}
\caption{To validate the random walk model, we did a Monte Carlo simulation integrating $10^5$ holes. The $x$-axis is horizontal distance from the starting point in mean free paths, $\lambda_h$. The blue histogram shows the spreading in the pn-junction plane after $10^6$ scattering events. The orange curve is the expectation from Equation~\ref{eq:psf}. Although the model is not perfect, about 92\% of the holes fall under the orange curve. We might furthermore expect the model to overestimate charge spreading at small radii while underestimating it somewhat at large radii. These conclusions are independent of the number of scattering events. We do not show $r$ in physical units because $\lambda_h$ has not been measured for a \WFIRST H4RG.}\label{fig:monte_carlo_histo}
\end{center}
\end{figure}

Writing in Nature, Karl Pearson posed a related problem in 1905.\cite{Pearson1905}
\begin{quote}
    {\it A man starts from a point O and walks $\ell$ yards in a straight line; he then turns through any angle whatever and walks another $\ell$ yards in a straight line. He repeats this process $n$ times. I require the probability that after these $n$ stretches he is at a distance between $r$ and $r+\delta r$ from his starting point, O.}
\end{quote}
Although Pearson's problem neglected drift and required uniform size steps, we have used Monte Carlo simulation (Figure~\ref{fig:monte_carlo_histo}) to show that it nevertheless usefully approximates charge spreading in HxRGs.

Rayleigh answered Pearson in the same volume,\cite{Rayleigh1905} pointing out that the solution was the same as that to an acoustics problem that Rayleigh had previously solved,
\begin{equation}
P_n(r)=\frac{2}{n\ell^2}\exp\left(-\frac{r^2}{n\ell^2}\right) r dr,\label{eq:rayleigh}
\end{equation}
with $r$ measured in the same units as $\ell$.

Of course, mobile holes in an HxRG do not undergo a two dimensional random walk. They move in three dimensions. Moreover, hole drift in response to \Edrift eventually overcomes the hole's thermal motion to sweep holes  out of the absorber and into the pn-junctions after $\approx n$ collisions, where they subsequently recombine. To account for these effects, we generalize Rayleigh's solution by replacing $\ell$ with the mean free path of holes, $\lambda_h$, projected onto the $xy$ plane,
\begin{equation}\label{eq:ell}
\ell=\frac{2}{\pi} \int_0^{\frac{\pi }{2}} \sin (\phi ) \, d\phi \lambda_h = \frac{2}{\pi}\lambda_h.
\end{equation}

With these substitutions, Eq.~\ref{eq:rayleigh} can be used to compute the charge spreading PSF,
\begin{equation}
\Psi\left(r\right)=\frac{\pi}{4n\lambda_h^2}\exp\left(-\frac{\pi^2r^2}{4n\lambda_h^2}\right).\label{eq:psf}
\end{equation}
By inspection, Eq.~\ref{eq:psf} is a two dimensional Gaussian with standard deviation,
\begin{equation}
\sigma=\pi^{-1}\lambda_h\sqrt{2n}.\label{eq:psi}
\end{equation}

We would like an expression for $\sigma$ in terms of the H4RG design parameters, operational parameters, and physical constants. The detector has been designed so that the drift field dominates diffusion in the vertical direction. We can therefore write an expression for the time needed for a hole to drift out of the absorber,
\begin{equation}
    \tau_d=\frac{z}{\mu_h \left|\bm{E}_d\right|},\label{eq:tau_drift}
\end{equation}
while the mean scattering time is,
\begin{equation}
    \tau_s=\frac{\mu_h m_h^*}{q}.\label{eq:scattering_time}
\end{equation}
Combining Eqns.~\ref{eq:tau_drift} and \ref{eq:scattering_time} yields,
\begin{equation}
    n=\frac{\tau_d}{\tau_s}.\label{eq:n}
\end{equation}

We can solve for the mean thermal speed of holes by equating the kinetic and thermal energy, yielding
\begin{equation}
v_{th}=\sqrt{\frac{k T}{m_h^*}}.\label{eq:hole_speed}
\end{equation}
Combining Eqns.~\ref{eq:hole_speed} and \ref{eq:scattering_time} provides an expression for the mean free path of holes,
\begin{equation}
    \lambda_h = v_{th}\tau_s.\label{eq:lambda_h}
\end{equation}

By substituting Eqns.~\ref{eq:n} and \ref{eq:lambda_h} into Eq.~\ref{eq:psi}, the charge spreading parameter can be rewritten as a function of the H4RG design parameters and temperature,
\begin{equation}
\sigma=\frac{1}{\pi}\sqrt{\frac{2 k T z}{q \left|\bm{E}_d\right| }}.\label{eq:stdev}
\end{equation}
Here $k$ is Boltzmann's constant, $T$ is temperature, and $z$ is vertical distance travelled (see Figure~\ref{fig:photodiode_architecture}).

One reader noted that the situation is similar to that in a fully-depleted CCD, which also has a drift field. In the case of the CCD, the drift field is an applied electric field created by the control voltages. Our Eq.~\ref{eq:stdev} is identical Fairfield~\etal's Eq.~2, except for our leading factor of $\pi^{-1}$.

Although more work is needed, the difference appears to relate to how motion in the $z$ dimension is treated. Fairfield~\etal refers back to Holland~\etal's Eq.~4.\cite{Holland1997} In their paper, Holland~\etal modelled charge spreading as purely two dimensional diffusion, with the duration for spreading set by the time required for a charge to drift out of the absorber.

We have treated the problem as a 2-dimensional random walk, but with the step size equal to the 3-dimensional step size projected onto the $xy$ plane, and with the duration equal to the same crossing time. There is a factor of $\pi^{-1}$ that appears when one projects the 3-dimensional step size onto the $xy$ plane (see Eq.~\ref{eq:ell}). In practice, our approach treats the vertical direction as a blend of drift and diffusion, but with drift clearly dominant. Fairfield~\etal treats the $z$-dimension as pure drift, which we believe is unphysical.

In any event, the two approaches are complementary. Both agree that: (1) charge spreading in these detectors should be well approximated by a Gaussian PSF and (2) the standard deviation of that PSF is given by Eq.~\ref{eq:stdev} to within a multiplicative constant of order unity. A quick check of crosstalk to the four nearest neighbors in one \WFIRST H4RG suggests that the leading factor of $\pi^{-1}$  that is used here is probably needed for \WFIRST H4RGs. Likewise, our Monte Carlo simulations support the leading factor of $\pi^{-1}$ (See Figure~\ref{fig:monte_carlo_histo}). We look forward to learning more as we test additional H4RGs.

Returning to Eq.~\ref{eq:stdev}, another reader commented that in the limit of small \Edrift, $\sigma$ becomes infinite: $\lim_{\left|\bm{E}_d\right| \to 0} \sigma = \infty$. In practice, this situation cannot arise. The derivation explicitly assumes that $\left|\bm{E}_d\right|$ is strong enough to eventually dominate the hole's thermal motion and sweep it into depleted \HgCdTeS. For detectors that lack the built-in drift field, like the old LPE NICMOS arrays, Eq.~\ref{eq:stdev} cannot be used. However, as we noted before, others have solved the diffusion equation that would govern charge collection in this case.\cite{Kirkpatrick1979,Holloway1986,Pickel2002}

Interestingly, hole mobility divides out and the \HgCdTeL $x$ parameter does not enter. This is helpful because we are not aware of any recent measurements of hole mobility in modern, high quality \HgCdTeS. The value shown in Table~\ref{tab:hgcdte-params} is an approximation that was arrived at by assuming that $\mu_h\approx 10^{-2}\mu_e$. This follows a common practice in the materials science literature, although a measured value would clearly be preferable.

Equation~\ref{eq:stdev} provides several useful insights. For example, if a detector already exists, then the only parameter that can be adjusted to improve charge diffusion is temperature, $T$. But, if one is developing new detectors, then $\left|\bm{E}_d\right|$ can be designed to reduce charge diffusion. Although one could also seek to reduce $z$ by using a thinner detector layer, in practice the detector's thickness is chosen to achieve acceptable QE at the cutoff wavelength.

\subsubsection{Observable Effects of Charge Diffusion}\label{sec:theory:spreading:practice}

In practice, a charge diffusion PSF having $\sigma_d\leq \frac{1}{5}\times$ the pixel pitch will be small compared to the optical PSF. Moreover, because Teledyne tightly controls both the thickness of the \HgCdTeS layer and the strength of \Edrift, we expect the charge diffusion PSF to vary little with position. However, it is a systematic effect.

The effects of charge diffusion will be somewhat more pronounced for cosmic rays. Because cosmic rays are not blurred by the optics, the dominant spreading mechanisms will be IPC, BFE, and charge diffusion (not necessarily in this order). When charge is deposited near the centers of pixels, IPC will often dominate because the other effects are small compared to the pixel size. However, when charge is deposited near pixel boundaries, charge diffusion and BFE will play larger roles.

\subsection{Brighter-Fatter Effect}\label{sec:theory:BFE}

Upon reaching depleted \HgCdTeS near the pn-junctions, charge diffusion stops and the motion of mobile holes is governed by the much stronger electric fields that exists there. BFE is a result of changes in the electrical field geometry in the depletion regions as charges integrate.

BFE is a nonlinear effect that was first described for visible wavelength CCDs\cite{Antilogus2014} and has since been observed in HxRG arrays.\cite{Plazas_2018a,Choi2020} It causes brighter point sources to produce wider images (as measured by the full width at half maximum) than fainter ones. As each new charge arrives, it experiences the combined electric field of the detector and the field produced by all previously integrated charges. In any small neighborhood of pixels, newly arrived charges are preferentially attracted to less full pixels and preferentially repelled from more full ones.

Figure~\ref{fig:photodiode_architecture} shows how these ideas can be applied to an HxRG array. Pixels 1 and 3 are close to reset and nearly empty, whereas pixel 2 has integrated more charge.  Recall that pixels integrate down, so the appearance of less charge in pixel 2 means that it has been exposed to more light. The volume of depleted \HgCdTeS that is associated with each pixel scales inversely with the integrated charge. More full pixels, like pixel 2, have less depleted \HgCdTeS associated with them, and therefore present less collection area to incoming holes.

For CCDs, a heuristic model exists in which ``pixel boundaries'' incrementally shift as pixels fill up.\cite{Antilogus2014} A group at Ohio State University recently extended this concept to include \WFIRST's H4RGs.\cite{Hirata2020} They validated their work using simulations\cite{Hirata2020} and H4RG test data acquired in the NASA Goddard Space Flight Center Detector Characterization Laboratory (DCL). Using flatfield data from one \WFIRST H4RG, they concluded the effective area of a pixel is increased by $(2.87\pm 0.03)\times 10^{-7}$ (stat.) for every electron in the 4 nearest neighbors, with a rapid $\sim r^{-5.6\pm 0.2}$ fall-off for more distant neighbors.\cite{Choi2020}

\subsection{Classical Non-linearity}\label{sec:classical-linearity}

HxRG's do not respond in a completely linear way to light. CNL is that part of the deviation from linearity that would be seen even with ideal, defect-free, and perfectly independent \HgCdTeS photodiodes. It is an idealization that depends only on the total integrated signal.

CNL originates at multiple places in the detector system. These include near the pn-junctions in the photodiodes. But, it also originates after the detector layer in the ROIC's source-follower amplifiers and other amplifiers in the readout electronics. Irrespective of where and when CNL originates, conventional astronomical pipelines treat it as a deterministic effect that can be removed by ``linearizing'' data\cite{Hilbert-2014} before fitting a straight line (in the case of up-the-ramp sampled data).

``Linearization'' is typically done using a set of training flats (spanning the full intensity range of interest) to measure a set of (typically 4) polynomial coefficients that fit the up-the-ramp samples. The fitted coefficients are then used to compute a multiplicative correction that maps each sample's value to what it would have been if the system had followed the same linear trend as the early, low signal, samples.

Our focus in this paper is on the underlying mechanisms that affect linearity. For those who are interested in calibrating data, or in simulating observations that include non-linearity as it is understood today, Hilbert\cite{Hilbert-2014} provides a good introduction in the context of HST WFC3 IR.

Most of the CNL probably originates within about $1-2~\mu$m of the pn-junctions (Figure~\ref{fig:photodiode_architecture}). Each pixel can be modeled as a diode in parallel with a non-ideal capacitor, $\rm C_{pix}$. The capacitor is non-ideal in the sense that $\rm C_{pix}$ depends on how much charge is stored. The pixel capacitance converts the integrated charge into the voltage that we measure.

$\rm C_{pix}$ is dominated by the cloud of charges surrounding the depletion region. As charge integrates, the depletion region collapses, and $\rm C_{pix}$ increases as a function of the changing geometry (smaller area; thinner depletion width). As $\rm C_{pix}$ increases, the number $\mu{\rm V}/h^+$ decreases in accordance with the capacitor equation.  

When the number of $\mu{\rm V}/h^+$ is measured at the SCA's output, it is known as the transimpedance gain, $g_t$. The transimpedance gain is an important parameter for designing detector readout systems. However, most astronomy instrumentalists are more familiar with the conversion gain, $g_c~\left(e^-/\rm DN\right)$, that is measured at the analog-to-digital converter output.

The conversion gain gives the conversion between digital numbers (DN) and integrated charge. The transimpedance and conversion gains encapsulate somewhat different CNL effects: $g_t$ includes only that which originates in the detector layer and ROIC whereas $g_c$ includes all of the CNL in the system.

Another important contributor to CNL is the MOSFET source-follower amplifiers in the HxRG ROIC, and other amplifiers in the readout electronics. Loose\cite{Loose:2003vh} provides a good overview of the basic HxRG ROIC architecture. Rauscher\cite{Rauscher2015} describes some of the noise mechanisms that operate. In any real electronic device, these are not perfectly linear. In modern HxRGs, our experience has been that the amplifiers' ~few percent deviation from linearity tends to be small compared to that caused by changing pixel capacitance. Although CNL enters at several points in the signal chain, in most cases we believe that the changing pixel capacitance is dominant.

A lot of excellent astronomy has been done by treating CNL as a deterministic effect separable from persistence, burn-in, and CRNL \etc For \WFIRST astrophysics, we believe that a more nuanced view might ultimately be required. Because so much of the CNL originates in the pn-junctions, we would not be surprised if charge trapping and release were to correlate CNL with other effects when ultimate control of detector systematics is required. This is something that we plan to work on going forward.

\subsection{Charge Capture and Release: Persistence, Burn-in, and CRNL}\label{sec:theory:traps}

Charge traps cause many of the most troublesome detector artifacts. These include persistence, burn-in, and CRNL. CRNL is also known as reciprocity failure and ``flux dependent non-linearity''. Although it is clear that there are correlations between these effects (see $\S$~\ref{det-per}), and probably also correlations with CNL, the detailed interrelations are not well understood today.

Charge traps are electrically active defect states in the semiconductor. The band theory of semiconductors assumes an infinite, periodic potential. However, real semiconductors are finite and have defects. Traps can capture charge; and also liberate charge by thermal stimulation or quantum tunneling, depending on the trap. In real semiconductors, the band theory therefore functions as a useful approximation, but for precision work one must allow for the effects of traps. How a given trap manifests depends on where it is in the semiconductor, both in terms of physical location and energy relative to the band edges.

There is an important difference between CCDs and IR arrays to keep in mind. In a CCD, one physically moves charge packets between pixels and senses them at the output. In an IR array, charge is sensed in place, without ever moving between pixels. One consequence is that in a CCD, a trap causes signal to vanish at some point in time, only to reappear later. We only see it when it shows up in the output amplifier after shifting down a column and then out a serial register. In an IR array, the time evolution is very different. When a charge becomes trapped, it falls part-way across the depletion region and gets stuck. In a sense, it partially integrates and we sense a partial change in the output voltage. When the trap subsequently releases, the previously trapped charge completes its journey, and we again see a partial change in the voltage.

The voltage change, $\delta V$, corresponding to trap capture or release very near the pn-junction is usually different from that for photo-generated holes, $\Delta V$, coming in from the absorber layer. This idea has been around since at least 2008, when it was presented as one of the pieces of evidence for Smith's ``theory for image persistence''.\cite{Smith-2008} Section~2 of Tulloch and George (2019)\cite{Tulloch2019} provides an excellent explanation, also in the context of persistence. They also report successfully measuring effective fractional charge. This is consistent with persistence (and other effects caused by trapping in the depletion regions) having different conversion gain than integrated photo-charge.

Intuitively, the idea that trapped charges integrate with different conversion gains makes sense. When a photo-generated hole is trapped, it partially integrates --and we see a smaller than usual change at the output, $\delta V_1$. When it is subsequently released by the trap (perhaps as persistence in the next exposure), it completes integrating and produces a change $\delta V_2$. Because $\rm C_{pix}$ is itself a function of the current charge state of the (non-ideal) pixel capacitance, it is likely that the sum of the two partial integrations does not equal the voltage drop produced by a photo-hole that integrates as the same time as the persistent hole: $\delta V_1 + \delta V_2\ne \Delta V$. We are still working to understand these inconvenient concepts, and on how to include them in calibration pipelines and simulated data. For now, in the interest of practicality, we have tried to provide guidance on how to do things as consistently as possible with today's state-of-the-art that largely ignores these complications.

Figure~\ref{fig:photodiode_architecture} shows several possible trap locations in modern HxRG detectors. Traps can occur anywhere, but here we have chosen to highlight material interfaces, where one would expect the defect density to be higher than in the bulk. Deep in the \HgCdTeS and near the pn-junctions, per $\S~\ref{sec:theory:absorption}$ we expect photo-generated holes to vastly outnumber photo-electrons (The main source of mobile electrons deep in the \HgCdTeS will probably be ionizing radiation). Table~\ref{tab:trap-descriptions} describes how we expect these traps to manifest in astronomical data. As we continue testing \WFIRST detectors, we will have an opportunity to test these hypotheses.

\begin{table}[htbp]
  \scriptsize
  \centering
  \caption{Charge Trap Descriptions}
    \begin{tabular}{ccp{14.835em}p{14.5em}p{14.585em}}
    \hline
    \hline
    ID &
      \multicolumn{1}{c}{Type} &
      \multicolumn{1}{c}{Description} &
      \multicolumn{1}{c}{Capture} &
      \multicolumn{1}{c}{Release}\\
    
    \hline
    $\alpha$ &
      $h^+$ &
      Hole trap near pn-junction,  HgCdTe:n. This category includes hole traps in the passivation that are close to the pn-junction. The trap is close to the junction and sometimes in depletion and sometimes not depending on how full the pixel is. &
      Loss of QE and non-linearity. Charge capture is very inefficient when in depletion, but may be efficient when out of depletion. Charge capture is most efficient when the detector is saturated because the depletion width is minimal then. &
      \underline{\bf Persistence:} Expected to integrate with $\delta V<\Delta V$ if in depletion. If out of depletion, integrates normally with $\delta V=\Delta V$.\\
    $\alpha$ &
      $e^-$ &
      Electron trap near pn-junction, HgCdTe:n,  . This category includes electron traps in the passivation that are close to the pn-junction. The trap is close to the junction and sometimes in depletion and sometimes not depending on how full the pixel is. &
      No immediate effect, but capable of causing persistence later. Capture is very inefficient when in depletion, but may be efficient when out of depletion. Capture is most efficient when the detector is saturated. &
      {\bf Persistence} if in depletion. Expected to integrate with $\delta V\lesssim\Delta V/2$.
      \\
    $\beta$ &
      $h^+$ &
      Hole trap near pn-junction, HgCdTe:p. This category includes hole traps in the passivation that are close to the pn-junction. The trap is close to the junction and sometimes in depletion and sometimes not depending on how full the pixel is. &
      Loss of QE and non-linearity. Capture is very inefficient when in depletion, but may be efficient when out of depletion. Caputure is most efficient when the detector is saturated. &
      {\bf Persistence,} Expected to Integrate with $\delta V\lesssim\Delta V/2$.
      \\
    $\beta$ &
      $e^-$ &
      Electron trap near pn-junction, HgCdTe:p. This category includes electron traps in the passivation that are close to the pn-junction. The trap is close to the junction and sometimes in depletion and sometimes not depending on how full the pixel is. &
      Little chance of capture. The material is p-doped. There should be almost no photogenerated electrons this deep in the HgCdTe. Thermally generated electrons are frozen out at these temperatures. &
      {\bf Probably undetectable.}
      \\
    $\delta$ &
      $h^+$ &
      Hole trap in the buffer-absorber interface &
      Loss of QE &
      {\bf Should be rare.} Persistence. Integrates like a hole created by a very blue photon. Normal $\delta V$.
      \\
    $\delta$ &
      $e^-$ &
      Electron trap in the buffer-absorber interface &
      No observable effect. &
      {\bf Probably undetectable.} The electron equilibrates with DSUB bias voltage.\\
    \hline
    \end{tabular}
  \label{tab:trap-descriptions}
\end{table}

\subsubsection{Persistence}\label{sec:theory:traps:persistence}

Persistence is a familiar but incompletely understood charge trapping phenomenon in IR arrays. It appears as a memory effect, whereby charge trapped during earlier exposures is released into the current one. Persistence is most obvious after saturating exposures, although at some level it probably happens for all exposures other than darks.

Smith~\etal (2008)\cite{Smith-2008,Smith-cali-2008} were the first to highlight the causative role of traps near the pn-junctions, and the interplay between charge traps the changing size of the depletion region. In 2013, Anderson and Regan\cite{Anderson2013} used electronic stimulus of a \JWST H2RG to mimic persistence and showed that the traps are not uniformly distributed. Rather, they appear to be clustered in a thin layer. This is consistent with the hypothesis that the pn-junction itself is where the electrically active defect density is highest.

Building on these ideas, and using lab results from three ESO H2RG detectors (both \swir and \mwir cutoff), Tulloch and George (2019)\cite{Tulloch2019} developed a model that they claim, ``describes the behavior of the trapped charges at a level adequate for useful correction.'' We note, however, that Tulloch and George used short LED pulses for illumination.

Previous H2RG testing for \JWST by Teledyne and the Goddard DCL, showed that quickly pulsing an LED gave different persistence from continuous illumination. Likewise; Long, Baggett, and MacKenty (2015; LBM)\cite{Long2015} used \Hubble WFC3 observations of the Omega Centauri and 47 Tucanae globular clusters to conclude, ``at a given fluence level, the power law index is shallower for longer exposures'' --indicating worse persistence for longer exposures. For two exposures to the same fluence = flux $\times$ time, the general consensus is that low flux and long integration time produces worse persistence than high flux and short exposure time.

We are not aware of any physical persistence model that is fully satisfactory for calibrating space astronomy data. One common heuristic is that employed by Tulloch and George, who phenomenologically modeled the trapped charge in each pixel, $Q\left(t\right)$, using a linear combination of exponentials,
\begin{equation}
    Q\left(t\right)=\sum_{i=1}^5 N_i . \left[1-\exp\left(\frac{-t}{\tau_i}\right)\right].\label{eq:tulloch23}
\end{equation}
In Eq.~\ref{eq:tulloch23}, $\tau_i\in\left\{10^0,10^1,10^2,10^3,10^4\right\}$ seconds is a set of ``bins'' representing different trapping time constants. The $N_i$ are a set of five numbers representing maximum charge trapped in a pixel for each of the $\tau_i$ bins. Not all groups use five coefficients, and it is common to see less.

LBM is the best empirical model that we are aware of for space astronomy. As has already been mentioned, LBM used WFC3 observations of Omega Centauri and 47 Tucanae to develop their model. The LBM model is built into the WFC3 calibration pipeline, although we believe that it is optional and not currently enabled by default. LBM did not seek to model the behavior of individual pixels. Rather, the same fit parameters were used for all pixels.

The LBM model uses three inputs to predict persistence. These are: (1) the fluence level in the earlier exposure, (2) the time since that exposure, and (3) the exposure time of the earlier exposure. The third input captures the idea that long and faint exposures have worse persistence than short and bright ones to the same fluence. LBM studied two persistence parameterizations and concluded that an ``exposure time dependent power law model'' best represented the data. It takes the form,
\begin{equation}
    P=A\left(\frac{t}{1000~s}\right)^{-\gamma}.\label{eq:LBM2}
\end{equation}
In practice, LBM fitted Eq.~\ref{eq:LBM2} to globular cluster data with $A$ and $\gamma$ being functions of fluence and then interpolated on the basis of the exposure time to estimate the persistence in other observations. For simulating \WFIRST observations, we recommend starting with Eq.~\ref{eq:LBM2}, LBM's ``exposure time dependent power law model''.

There has been a lot of good work on what causes persistence recently. For the reader who desires more information, we recommend starting with Tulloch and George\cite{Tulloch2019}, and following the chain of references back from there. One minor issue to be aware of is that most papers show n-on-p photodiodes, with the polarity reversed compared to our Figure~\ref{fig:photodiode_architecture}. We think this is benign, and the arguments are equally valid irrespective of the polarity of the charges that the authors choose to discuss. 

\begin{figure}[t]
    \centering
    \begin{subfigure}[t]{.48\textwidth}
        \centering
        \includegraphics[width=\textwidth]{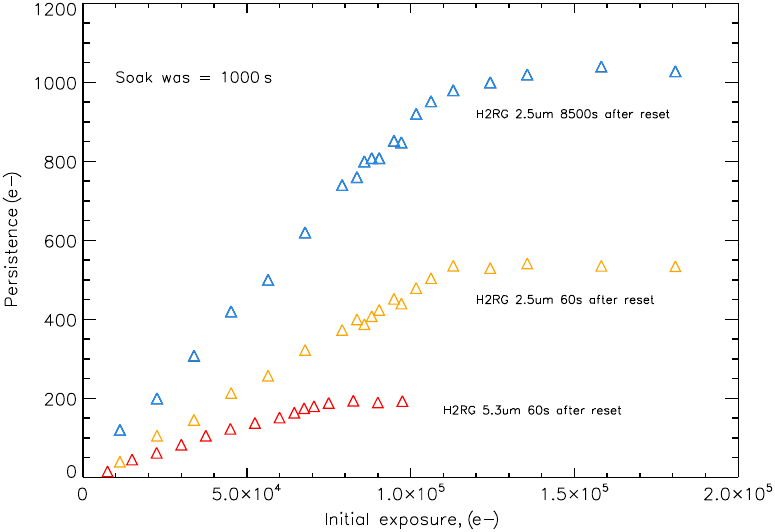} 
        \caption{Persistence increasing linearly}
    \end{subfigure}
    \begin{subfigure}[t]{0.48\textwidth}
        \centering
        \includegraphics[width=\textwidth]{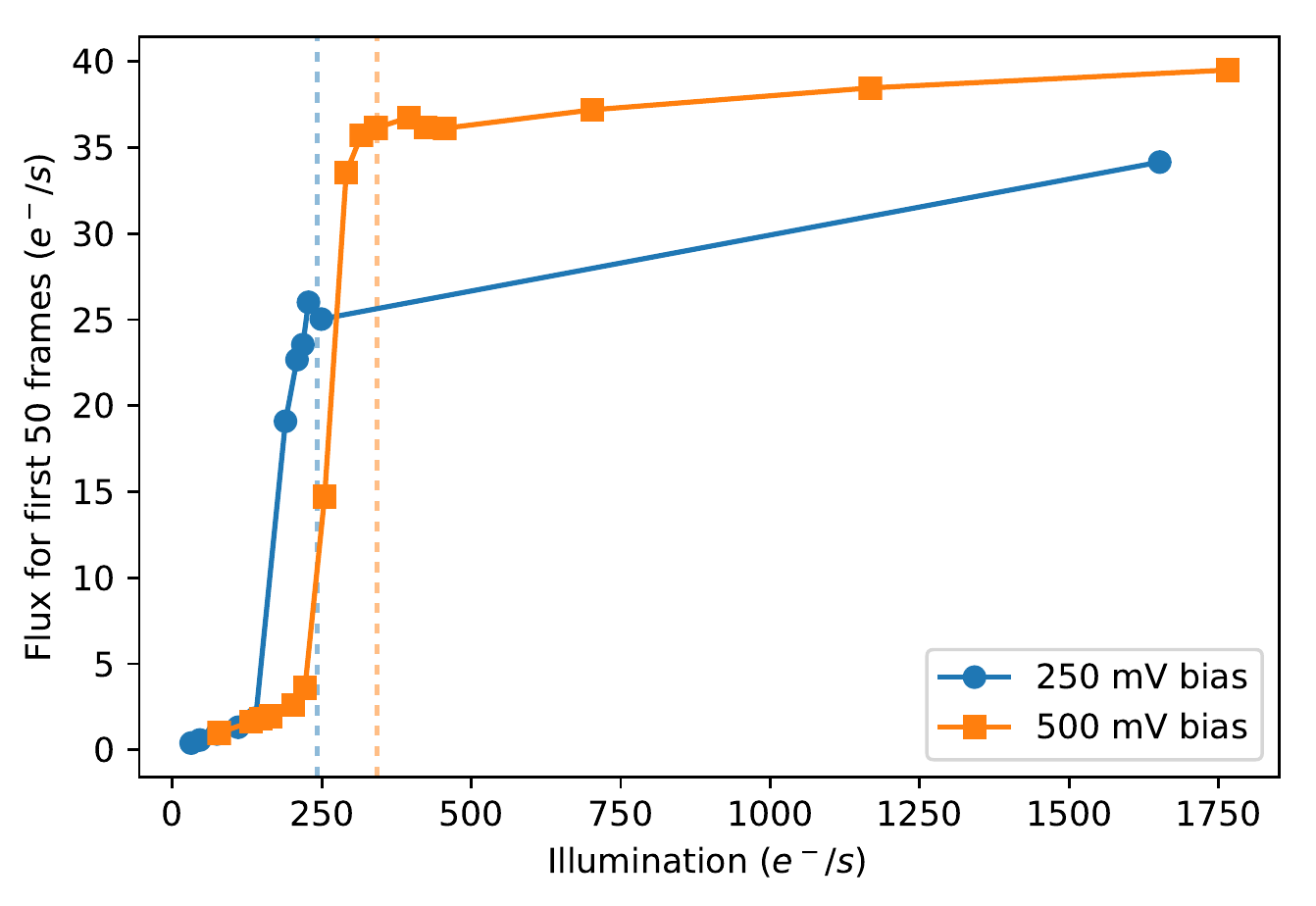} 
        \caption{Persistence threshold effect}
    \end{subfigure}
    \caption{Panel (a) shows the most ``desirable'' form of persistence. The persistent signal increases approximately linearly, before slowly rolling over and saturating at some level. Once saturated, it does not get any worse. Panel (b) shows a particularly undesirable form of persistence that was first observed in some Euclid prototype detectors. Persistence increases approximately linearly, before increasing dramatically near detector saturation. The persistence subsequently increases less rapidly, but remains at a much higher level than at lower signal levels. The \WFIRST Project spent considerable effort refining the H4RG design with the aim of eliminating the persistence threshold effect insofar as practical, but we may nevertheless see it in some of the \WFIRST H4RGs. Panel b replots data from Ref.~\citenum{Tulloch2019}.}\label{fig:persistence}
\end{figure}

How persistence manifests depends on the particular detector, its operating temperature, source brightness, exposure duration, and the characteristics of the traps. For example, the \JWST H2RGs (\swir and \mwir) are operated near $\rm T = 40~K$. For the \JWST detectors at this temperature, persistence increases linearly with fluence before saturating at some value that depends on the detector. The ESO Detector Group measured similar behavior (Figure~\ref{fig:persistence}a) in two \swir H2RGs operated at 80~K and one \mwir H2RG operated at 40~K.\cite{Tulloch2019} Although persistence is never desirable, this behavior is better than a thresholding effect that has been seen in some Euclid and \WFIRST HxRGs.

Figure~\ref{fig:persistence}b shows how persistence manifests in an HxRG that has the thresholding behavior. As before, persistence increases linearly for small signal. However, as full well is approached, persistence increases dramatically before leveling off somewhat at a level that is much higher than that seen at low signal. We speculate that the thresholding effect may relate to operating temperature because it was not seen to anywhere near the same degree in the \JWST production (\JWST's detectors operate colder than Euclid and \WFIRST).

In any event, the \WFIRST project realized before starting the flight production that this thresholding behavior was undesirable. The flight H4RGs include design modifications that are intended to reduce it, and the mission operation concept includes constraints that are intended to minimize the time that \WFIRST's H4RGs spend in saturation.

\subsubsection{Burn-in}\label{sec:theory:traps:burn-in}

Burn-in is persistence's less well known cousin. Regan (2012)\cite{Regan2012} describes the effect as ``inverse persistence'' and also relates it to CRNL. Whenever a pixel transitions from staring at faint to bright light, the pixel is less sensitive for some period of time while the traps equilibrate to the new flux. In the case of the prototype \WFIRST H4RGs that we have examined, a burn-in effect is detectable and time-dependent for the first $\approx 12$~minutes after a pixels transitions seeing dark to light. 

\begin{figure}[t]
    \centering
    \includegraphics[width=.8\textwidth]{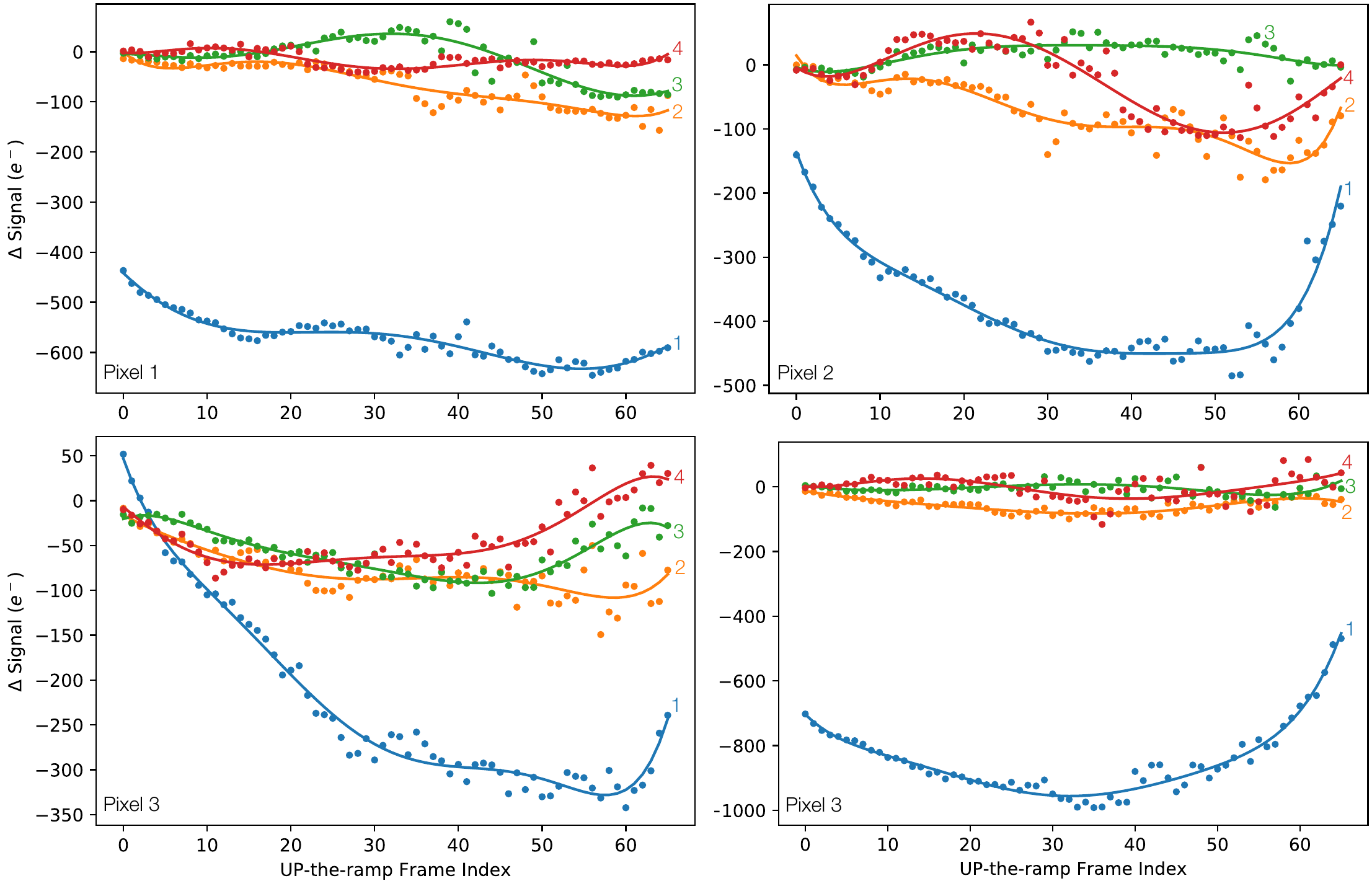}
    \caption{This figure shows the burn-in effect as it appeared in one prototype \WFIRST H4RG. Each panel shows a different pixel extracted from a flat field exposure to the onset of saturation. The solid lines are low-order polynomial fits to guide the eye. The x-axis values are the up-the-ramp (UTR) frame indices in each of four sequential 65 frame SUTR exposures ($\textrm{EXPTIME}\approx 182~s$). A preliminary analysis showed that the burn-in effect was usually undetectable after the $4^{th}$ exposure ($\approx$12 minutes) in a sequence. The y-axis shows the pixel value in the labeled exposure minus the mean of all $5^{th}$ and later exposures. We tentatively attribute the sharp upturns in some curves for $x\gtrapprox 60$ to saturation effects. There was considerable pixel-to-pixel variation in burn-in properties, although the general trend of burn-in being much less noticeable after the first exposure held constant.}
    \label{fig:burn-in}
\end{figure}

Figure~\ref{fig:burn-in} shows how the burn-in affect manifested in an exposure sequence that interleaved random numbers of darks and illuminated flats. The burn-in effect was clearly most detectable in the first exposure after changing from dark to bright. The upturns at high frame index in each plot are approximately coincident with the onset of saturation.

The \WFIRST team is not the first to see a burn-in effect. Regan\cite{Regan2012} characterized the effect in a \mwir cutoff \JWST H2RG. While it is plausible that the burn-in that we see in \WFIRST H4RGs might be inverse persistence, the sizes of the two effects are different when measured in DN. The DN that are lost while burning-in do not all reappear as persistence. One plausible explanation might be that the transimpedance gains, $g_t \left(\mu {\rm V}/h^+\right)$, for the two processes could be different with $g_{t,\rm burn-in}>g_{t,\rm persistence}$. In this case, the number of trapped holes could be preserved while still allowing for different signals in DN to be measured.

Compared to persistence, burn-in is a fairly new effect. We are not aware of any parameterizations in the literature yet. However, we do have some DCL test data that probably will eventually be used to study the effect further.

\subsubsection{Count Rate Dependent Non-Linearity (CRNL)}\label{sec:theory:traps:CRNL}

In addition to the well known CNL, HxRG arrays suffer from count rate dependent non-linearity (CRNL). For a detector with CRNL, the apparent QE depends on incident flux. The sense is such that the detector become more sensitive when exposed to brighter light.

Intuitively, any time the detector undergoes a significant change in illumination level or charge state, the traps in and around the pn-junctions will seek to equilibrate to the new condition. If the detector was previously looking at something much brighter, then there will be a $\sim$few minute interval during which persistence is obvious. If it was previously looking at something much fainter, then then will be a $\sim$few minute interval during which burn-in is obvious. After the persistence and burn-in have settled, or been avoided with operational constraints, CRNL is what remains. Like persistence and burn-in, we believe that CRNL originates in a thin layer of traps (the same layer of traps) located in and immediately surrounding the pn-junctions.

The current parameterization of CRNL (to be discussed shortly) is independent of time, although we can imagine that this might change as more is learned. In practice, we expect there to be a complex interplay between all effects that originate in and around the depletion regions; including persistence, burn-in, CRNL, and also perhaps CNL and dark current. For the highest fidelity simulations, we think that it will ultimately be important to simulate most of these effects in an integrated, time dependent framework that allows for correlations.

The CRNL in the older liquid phase epitaxy NICMOS detectors had a strong color dependence.\cite{DeJong2006} However, we believe that it is unlikely that similar color dependence will be seen in modern H4RGs. The reasons are: (1) a sapphire growth substrate was left on in the NICMOS arrays and (2) the NICMOS arrays had no drift field $\left(\bm{E}_d=\bm{0}\right)$. With reference to Figure~\ref{fig:photodiode_architecture}, no sapphire substrate is used in modern devices, but if it was, it would be located between layers a and b in this figure. These two factors mean that there was an abundance of traps near the Sapphire to \HgCdTeS interface and there was no strong electric field to drive mobile carriers away from the traps. One would therefore expect to see strong color dependence in the NICMOS detectors because the absorption depth depends on color as per Eq.~\ref{eq:beer-lambert}. The DCL is working to validate these hypotheses now as they have important implications for simplifying the \WFIRST Relative Calibration System (RCS).\cite{Wirth2019}

Among NASA missions, CRNL was first noticed in NICMOS.\cite{Bohlin2005,Bohlin2006,DeJong2006} De Jong\cite{DeJong2006} describes how the NICMOS flatfield lamps were used to characterize CRNL using a Lamp-On/Lamp-Off (LOLO) technique.
\begin{quote}
    {\it A star cluster field was imaged in a lamp off-on-off sequence in all cameras in a selected set of filters, followed by a series of darks to investigate persistence and to clean the images from any remaining charge for the next orbit. Subtracting the lamp-off images from the lamp-on images clearly shows residual ADUs at the star positions, indicating that a higher background (and thus total) count rate increases the number of ADUs registered from an object.}
\end{quote}
An ADU is an analog to digital converter unit. For \WFIRST, current plans call for the WFI to have a LOLO capability, although the implementation details are still being developed.

These results suggest that the detector was more sensitive for brighter targets. There was also the previously mentioned color dependence that we do not expect to see in modern HxRGs. The CRNL amplitude was reported to be $0.06-0.10~\rm mag$ offset per factor ten (``dex'') change in incident flux for the shortest wavelengths $\left(0.9~ {\rm and~} 1.1~\mu\rm m\right)$, about $0.03~\rm mag~ dex^{-1}$ at $1.6~\mu\rm m$, and less at longer wavelengths.

\begin{figure}[t]
    \centering
    \includegraphics[width=\textwidth]{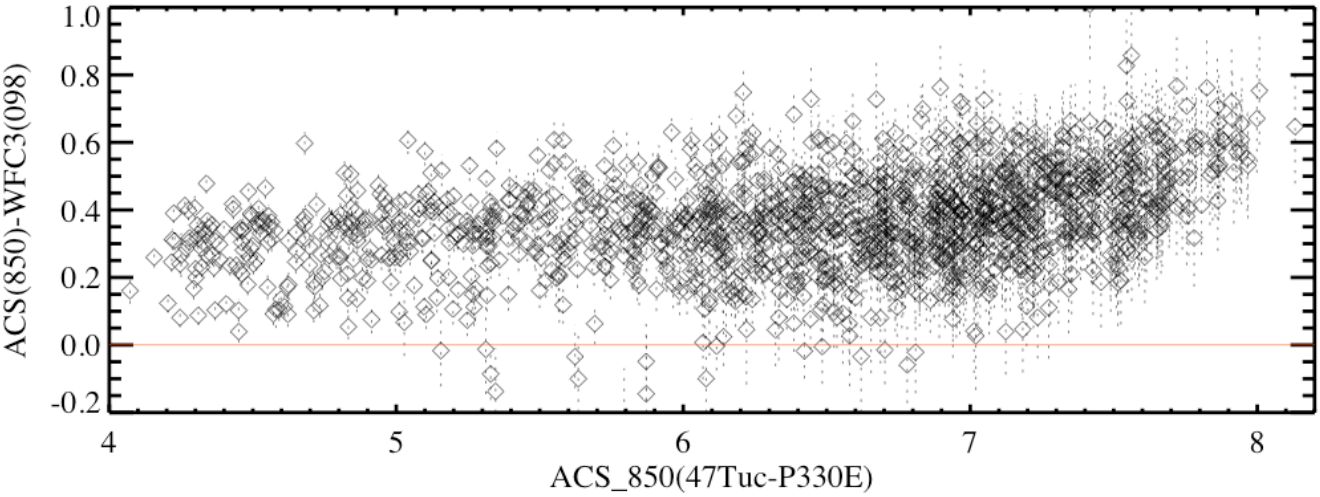} 
    \caption{a) CRNL manifests as a flux dependent change in apparent QE. The sense is such that the apparent QE is comparatively higher for brighter sources. This plot compares aperture photomery in the 47 Tuc globular cluster for a set of stars at a wavelength where the ACS and WFC3 filters overlap. Because the ACS CCDs are known to be highly linear, the change is attributed to CRNL in WFC3's H1R detector. The CRNL's amplitude is about $\rm 0.1\%~dex^{-1}$. P330E is an infrared standard star. The filter definitions are: $\lambda_{\rm ACS}=883-1006~\rm nm$ and $\lambda_{\rm WFC3}=908-1065~\rm nm$. This figure is reproduced from Ref.~\citenum{Riess2010}.}\label{fig:riess}
\end{figure}

As would be expected with an HxRG family detector, the CRNL evidenced by WFC3 IR's H1R detector is smaller and not color dependent (to within the uncertainties). Riess (2010)\cite{Riess2010} used stellar photometry in overlapping \Hubble ACS CCD and WFC3 IR passbands to measure CRNL = $0.010 \pm 0.0025~\rm mag~dex^{-1}$ independent of wavelength (Figure~\ref{fig:riess}). Using a more extensive set of astronomical calibrators; Riess, Narayan, and Calamida (2019)\cite{riess2019a} measured, ``CRNL in WFC3-IR to be $0.0077 \pm 0.0008~\rm mag~dex^{-1}$ (or $0.0075 \pm 0.006~\rm mag~dex^{-1}$ including [\ldots] grism measurements), characterized over 16 magnitudes with no apparent wavelength dependence [\ldots].''

For \WFIRST, we expect the CRNL properties to be similar to WFC3 IR, albeit somewhat better because Teledyne's process tends to improve with time. The DCL is working to validate this assumption now. For simulation purposes, a simple power law encapsulates today's understanding. Within the \WFIRST project, one common parameterization is the one described by Christopher M. Hirata at a Project meeting in February, 2019. Neglecting realization dependent drifts (temperature variations, electronic 1/f noise, \etc), it is,
\begin{equation}
    I_{\rm obs}=C I \left(\frac{I}{I_{\rm ref}}\right)^{\alpha_{\rm CRNL}-1},\label{eq:hirata}
\end{equation}
where $I$ is an intensity in units of photogenerated charge per second.

On physical grounds, we do not believe Eq.~\ref{eq:hirata} to be a complete description of CRNL. Like persistence and burn-in, it should depend on time, at least until the initial persistence and burn-in have settled down. Likewise, because CRNL orginates in the same part of the photodiodes that governs persistence, burn-in, dark current, and (much of the) CNL; we think it is likely that it will eventually be necessary to model correlations between CRNL and these other instrument signatures. However, Eq.~\ref{eq:hirata} reflects what is known today. As such, it is a reasonable starting point so long as one understands that it may need to be revised as we learn more.

\subsection{Inter-pixel Capacitance (IPC)}\label{sec:theory:IPC}

Inter-pixel capacitance (IPC) is an effect that originates in the indium interconnects and the electrical contacts on both sides of the interconnects (see Figure~\ref{fig:photodiode_architecture}; component h and nearby structure). Parasitic capacitance between these parts of the pixels correlates each pixel with nearby neighbors, causing a slight but significant blurring of the PSF. In practice, in \WFIRST H4RGs the 4 nearest neighbors see roughly 2\% IPC, and the next nearest-neighbors see roughly 0.2\% IPC. The effect becomes very small for more distant pixels, although it may still remain important for some very demanding observations.

IPC in modern astronomical array detectors was first described by Moore~\etal (2006)\cite{Moore-2006}. They introduced the concept of an IPC kernel that effectively blurs the scene. For \WFIRST, we are finding that IPC has some spatial dependence, so a single IPC kernel would not be appropriate for all pixels in an H4RG.

\subsection{Ordering of Effects in Simulations}\label{sec:theory:ordering}

\WFIRST is fortunate to have a passionate community of interested scientists. They are organized into Science Investigation Teams (SIT) that map onto the different science projects that \WFIRST is designed to implement. Many SIT members are interested in simulating observations. We are therefore often asked, in what order should one apply the various non-ideal behaviors?

Insofar as possible, we recommend applying the effects in the order in which they occur in nature. The effects in the depletion regions are correlated, time dependent, and happen in parallel. Mathematically, they do not commute---one can expect to get somewhat different answers depending on the ordering if applied sequentially. For the effects that originate in the pn-junctions, we believe that for the highest fidelity simulation, one should treat them in an integrated time-dependent framework---not as a commutative sequence of separable operations. In this regard, $\S~3.4$ of Hirata and Choi\cite{Hirata2020} provides an interesting mathematical way of viewing the interdependence of IPC and CNL.

However, we understand that we are all learning, and that it may be desirable to start with simple parametric models and build up to higher fidelity time dependent models of the physical effects. If a simple parametric model is desired, then it is not clear that there is any preferred order for applying the effects that happen in the photodiodes. Our recommendation is to try different combinations, and compare the resulting simulations to real data to find the ordering that appears most realistic.

This paper is entirely about \WFIRST's H4RG detectors. In the near future, we will start to learn more about the ACADIA application specific integrated circuits (ASIC) that will be used to control the H4RGs\cite{Loose-2018}. For now, our recommendations cover non-ideal behaviors that originate in the detectors only. As we learn more about the ACADIAs, we look forward to saying more.

\begin{figure}[t]
    \centering
    \includegraphics{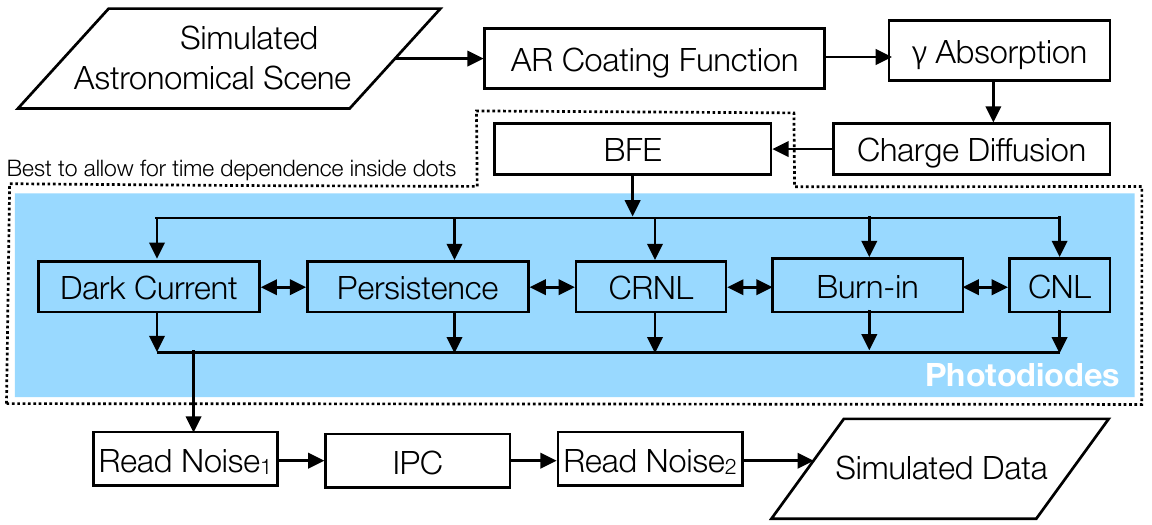}
    \caption{This figure shows the recommended sequence for simulating the effects that we have described. Starting with BFE and ending just before Read Noise$_1$, the effects happen in parallel and there are time-dependent correlations. For the highest fidelity simulation, we recommend simulating the effects in the dotted region in an integrated framework that allows for the time dependence and correlations. Read noise$_1$ may be modeled as (white) Johnson-Nyquist noise arising in \HgCdTeS contact resistance. For \WFIRST, its amplitude will be roughly 10~$e^-/\rm read$. Read Noise$_2$ originates in the ROIC and readout electronics. See Rauscher (2015)\cite{Rauscher2015} for a discussion of electronic read noise mechanisms in HxRG systems.}
    \label{fig:order_of_effects}
\end{figure}

With these caveats in mind, Figure~\ref{fig:order_of_effects} shows the recommended sequence for simulation.

\section{Measured Detector Performance} \label{det-per}

The Goddard Detector Characterization Lab (DCL) has characterized banded array lot, full array lot, yield demonstration lot, and has started characterizing flight lot detectors for WFIRST. The DCL at NASA Goddard Space Flight Center is an engineering team of detector scientists and engineers that focus on characterizing the optical and electrical performance of large format detector arrays at wavelengths from the ultraviolet to the infrared. The DCL hosts under one roof laboratories and clean rooms to perform sensitive measurements of detector properties such as readout and total noise, relative and absolute quantum efficiency, the modulation transfer function, crosstalk and interpixel capacitance, linearity and full well depth, as well as image persistence to name a few key capabilities. The DCL also supports flight qualification testing including radiation, thermal vacuum, and vibration testing. All the data presented in this paper was collected and analyzed by Goddard's DCL.

We present baseline measurements of WFIRST H4RG-10 detector properties such as gain, noise and dark current below. We also present detector persistence, linearity, quantum efficiency and interpixel capacitance performance. All tests are performed using a Leach controller system from Astronomical Research Cameras, Inc. to readout a tested array in 32 channels at a pixel sampling rate of $\sim$200~kHz. All tests are performed with the detector at 100~K with the detectors operating at 1.0~V reverse bias unless noted differently.

The WFIRST H4RG-10s are nominally operated at 1.0~V reverse bias, providing a large dynamic range ($\sim$100~ke- full well) and low read noise for science observations.  This operational bias is higher than, for example, the operational bias of WFC3 and the \JWST detectors which operate at $\sim$ 0.25--0.5~V bias\cite{Rauscher_2014}. Early in the detector development process a trade was identified between read noise performance and dark current performance. It was found that at increased bias, read noise could be decreased with only marginal increases of dark current, which was nearly immeasurable.

In addition to the results from the aforementioned tests, we summarize results from a new suite of tests that have been developed to look at the BFE in select arrays. We discuss results of environmental testing on these arrays from the full array lot. Finally, we provide an update on current flight lot detector performance.

\subsection{Gain measurement}
The ADC conversion gain for a detector system is a fundamental property that describes how digitized counts from an ADC map to photo-generated electrons detected by the system. This calculated conversion gain is used to convert ADU or DN in an acquired image to photo-generated electrons. This gain is used to calculate subsequent detector performance parameters, such as noise and dark current, to physical units, i.e. e- and e-/s. As described in \S\ref{sec:theory}, our detectors actually detect photogenerated holes. However, we keep with the common convention of referring to the photogenerated charge in units of e-. 

The conversion gain is typically measured using a photon transfer curve methodology\cite{Janesick-2007,Mortara-1981}. In this method, a series of exposures of different exposure times are taken with a constant illumination source to generate a sample of data to plot the mean signal in each exposure against the variance of the signal in each exposure. This data is fit with a line, and following photon transfer curve theory, the inverse of the slope of this line is the ADC conversion gain. However, care must be taken before calculating the conversion gain to remove any interpixel capacitance in the data. The presence of interpixel capacitance has been shown to artificially decrease the variance in a measured signal, decreasing the slope of a mean-variance plot, and thus increasing the calculated gain\cite{Moore-2006,Fox-2009}. 

For the detector data presented below, the gain was measured using 10 exposures with 11 samples up-the-ramp (total exposure time $\sim$30~s, total signal $\sim$2000~e-) at constant flux using a uniform 1.4~$\mu$m light source. For each exposure, we calculate the global mean signal and variance in the signal across the entire array for each frame up-the-ramp. The inverse of the slope of a linear fit to these measurements produces an estimate of an average conversion gain. The median conversion gain is then taken across the 10 average conversion gains calculated from the exposures. This median conversion gain is used to convert from data units to physical units for the array performance properties below. This conversion gain is corrected for the IPC as measured in \S{\ref{sec:crosstalk}} using equations from Ref~\citenum{Moore-2006}. This correction factor is typically 0.8 -- 0.9 for the median IPC values of 1.5--1.9\%.

\subsection{Noise performance}
The precision of a detector system is characterized by the system's noise performance. Systems with better noise performance more efficiently detect signal and likewise have increased sensitivity. Some of the dominant sources of noise in the absence of illumination are located in the readout electronics (e.g. source follower MOSFETs), and we call the resultant noise from these sources the read noise. The read noise describes the typical variation of recorded counts from one read to the next in the limit of no illumination. Astronomers typically work with illuminated exposures, and in particular, are interested in object fluxes. A flux image (e.g. counts/s)---formed from an up-the-ramp group science exposure of a given integration time---will be subject to additional sources of detector noise. For example, for long exposures, shot noise from the dark current of the detector may become important for very faint signals or very high dark current. The noise performance will also depend on how the flux is estimated from the samples. To more fully describe the noise performance, we describe the WFIRST H4RG-10 detector noise performance for two representative sampling techniques: correlated double sampling (CDS)---when numerous samples are not possible and a correlated double sample is used; and line-fitting---when numerous samples of a signal are possible and a slope is calculated up-the-ramp. The noise performance for these estimations of detected signal are presented next.

\subsubsection{CDS noise measurement}
When source brightness readily saturates the detector it may not be possible to collect numerous samples of the signal up-the-ramp. In this regime, a correlated double sample (CDS) is often taken. The detector array is reset, then read out at least twice, generating two frames taken at least 1 frame time apart. The signal detected is estimated by the difference between the frames.

\begin{figure*}[h]
\begin{centering}
\begin{subfigure}[t]{0.45\textwidth}
        \centering
        \includegraphics[width=0.7\textwidth]{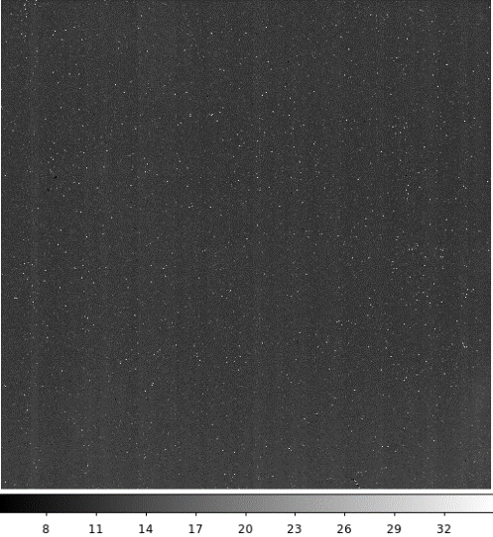} 
        \caption{CDS noise image for SCA 18237.} \label{fig:cdsnoise-img}
\end{subfigure}
\begin{subfigure}[t]{0.5\textwidth}
        \centering
        \includegraphics[width=\textwidth]{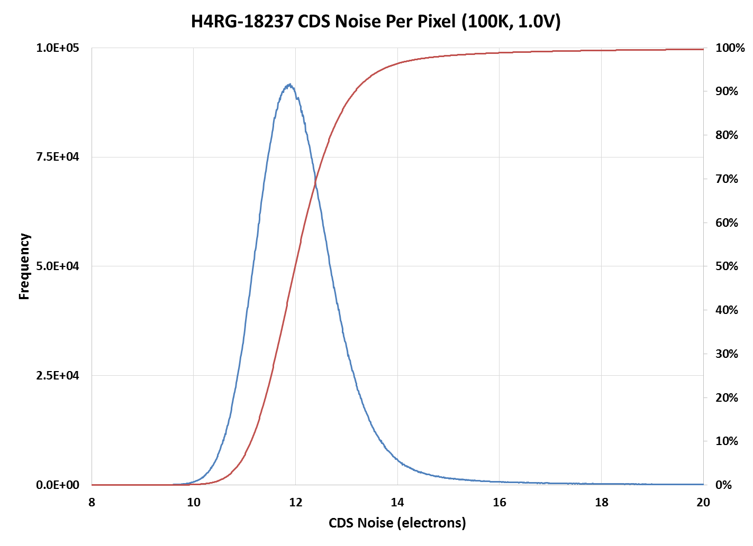} 
        \caption{CDS noise histogram for SCA 18237.} \label{fig:cdsnoise-hist}
    \end{subfigure}
\caption{(a) An example of a master CDS noise image for the H4RG-10 device 18237. (b) The histogram of the master CDS noise image. Using these images and histogram we can estimate the median and per pixel noise performance of the H4RG-10 devices. SCA 18237 has median CDS noise of $\sim$12~e-. }\label{fig:cdsnoise}
\end{centering}
\end{figure*}

To estimate the detector system noise in such a CDS measurement, five exposures of 100 consecutive samples up-the-ramp are acquired with no illumination. These frames are reference pixel corrected to account for bias drifts. Then for each exposure, the non-overlapping pairs of images in each exposure are subtracted to form 50 CDS frames. We take the standard deviation per pixel across these 50 frames and form a single exposure CDS noise frame. The median per pixel is then taken across these five noise frames to form the final master CDS noise frame. The noise statistics (i.e. mean, median, mode) are then calculated from this master image (see Figure~\ref{fig:cdsnoise}). All detectors produced in the full array lots have achieved acceptable CDS noise performance ($<$20~e-). See Table~\ref{tab:summary} for a summary of performance characteristics for these PV3 devices.

\subsubsection{Line fitting noise measurement}
For moderately faint signals it is often possible to take numerous samples of the the source signal up-the-ramp. In this regime, we implement a line-fitting routine to estimate the flux per pixel by calculating the slope per pixel of each pixel's signal up-the-ramp as a function of time (e.g., Refs.~\citenum{Chapman-1990,Garnett-1993}).

To estimate the noise in our detector system using this sampling technique, we acquire 100 exposures of 55 samples up-the-ramp with a total exposure time of $\sim$150~s. These frames are reference pixel corrected to account for bias drifts. Next, the slope is calculated per pixel for each exposure forming a slope image, and the standard deviation per pixel is calculated across the 100 slope images. The derived noise in the slope (flux) per pixel is then multiplied by the exposure time (150~s) and the estimated conversion gain to estimate the total noise in electrons. The median total noise performance of the PV3 passivation full array lot devices is $\sim$7~e-. See Table~\ref{tab:summary} for a summary of performance characteristics for PV3 devices.

\subsection{Dark current performance}
Dark current is the general term to describe the leakage current from the detector material (the HgCdTe photodiodes for WFIRST's H4RG-10s) into the detector system. The most common form of dark current is thermally generated dark current. As dark current originates in the semi-conductor material it can usually be controlled and minimized by cooling the detector to low temperatures. Typical dark currents for shortwave infrared HgCdTe detectors at cryogenic temperatures are sub-electron. But even at cryogenic temperatures, it is still important to quantify dark current as it adds a source of shot noise to acquired signals.

\begin{figure*}[h]
\begin{centering}
\begin{subfigure}[t]{0.45\textwidth}
        \centering
        \includegraphics[width=0.7\textwidth]{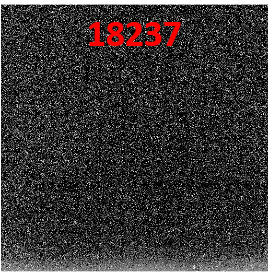} 
        \caption{Dark current image for SCA 18237.} \label{fig:sca-18237-dark-img}
\end{subfigure}
\begin{subfigure}[t]{0.45\textwidth}
        \centering
        \includegraphics[width=0.7\textwidth]{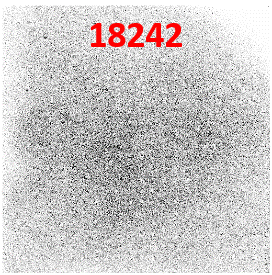} 
        \caption{Dark current image for SCA 18242.} \label{fig:sca-18242-dark-img}
    \end{subfigure}
\caption{(a) An example of the average dark current image for the H4RG-10 device 18237. (b) An example of the average dark current image for the H4RG-10 device 18242.  Using these images, we can estimate the median and per pixel dark current performance of the H4RG-10 devices. SCA 18237 has a nearly immeasurable median dark current of 0.001~e-/s. SCA 18242 has higher median dark current (0.04~e-/s) but still represents acceptable performance. }\label{fig:dark-img}
\end{centering}
\end{figure*}

To measure dark current we acquire at least 4 long exposures with 101 samples up-the-ramp and a total exposure time of $\sim$7258~s after the detector is cooled (100~K), un-illuminated and stable. These frames are reference pixel corrected to account for bias drifts. The first frame is then subtracted from subsequent frames for each exposure. Next, the slope (flux) per pixel is measured in each of these exposures forming four slope images. The median across these slope images is taken to generate a high signal-to-noise dark current map for the arrays. Figure \ref{fig:dark-img} shows the dark current images from two representative devices.

The median dark current across all PV3 full array lot devices is 0.002~e-/s, which is consistent with our estimated measurement error of 0.003~e-/s (the typical variation seen between separate dark current measurements of the same device). Many of the detectors have values below this measurement error value suggesting the dark is so low that it is imprecisely measured in the current setups. See Table~\ref{tab:summary} for a summary of performance characteristics.

\subsection{Persistence performance}
Persistence is the observation of released photo-generated charge from a previous readout in a subsequent readout. Persistence is theorized to originate from traps caused by defects in the detector material. For astronomers and scientists, persistence represents a potentially large source of background contamination if the amount of trapped charge released from a previous frame overwhelms a faint signal of interest. This spurious signal is not easily subtracted and depends on the complete illumination history of the detector making it a critical limitation for a detector. For these reasons, the WFIRST detector development program sought to select a pixel design that minimized persistence. This was accomplished with the PV3 passivation recipe identified from the banded array lot. We present the persistence performance of the full array lot with the PV3 passivation.

\begin{figure*}[h]
\begin{centering}
\begin{subfigure}[t]{\textwidth}
        \centering
        \includegraphics[width=\textwidth]{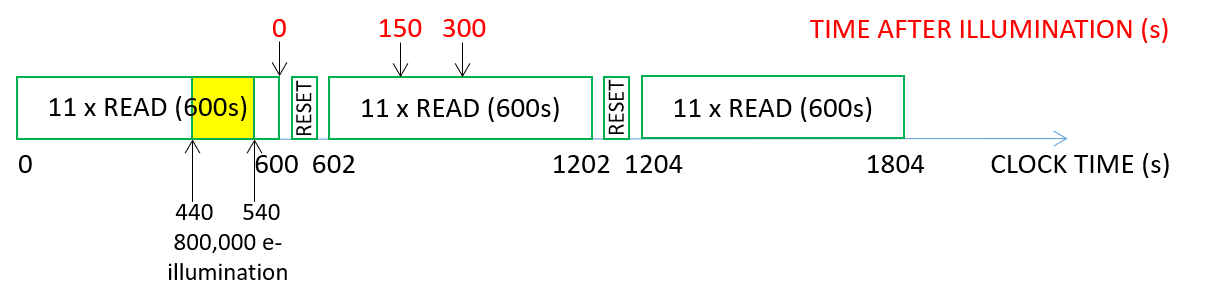} 
\end{subfigure}
\caption{Persistence data acquisition. The detector under test is first illuminated in a 600 second exposure with 11 samples up-the-ramp to a total signal level that is approximately 800~ke- or about eight times the typical full well for these devices. The amount of persistence is observed in subsequent identical exposures (600 s, 11 samples up-the-ramp) with no illumination.}\label{fig:per-acq}
\end{centering}
\end{figure*}

Persistence is measured in a dark exposure ramp following an initially illuminated ramp. For the full array lots presented, persistence is measured with the detector at 100~K using 1.4~$\mu$m illumination. The detector under test is first illuminated for 100 seconds during a 600 second exposure with 11 samples up-the-ramp to a total signal level that is approximately 800~ke- or about eight times the typical full well (FW) for these devices. The amount of persistence is measured in subsequent identical exposures (600 s, 11 samples up-the-ramp) with no illumination (see Figure~\ref{fig:per-acq} for acquisition schematic).

\begin{figure*}[h]
\begin{centering}
\begin{subfigure}[t]{\textwidth}
        \centering
        \includegraphics[width=\textwidth]{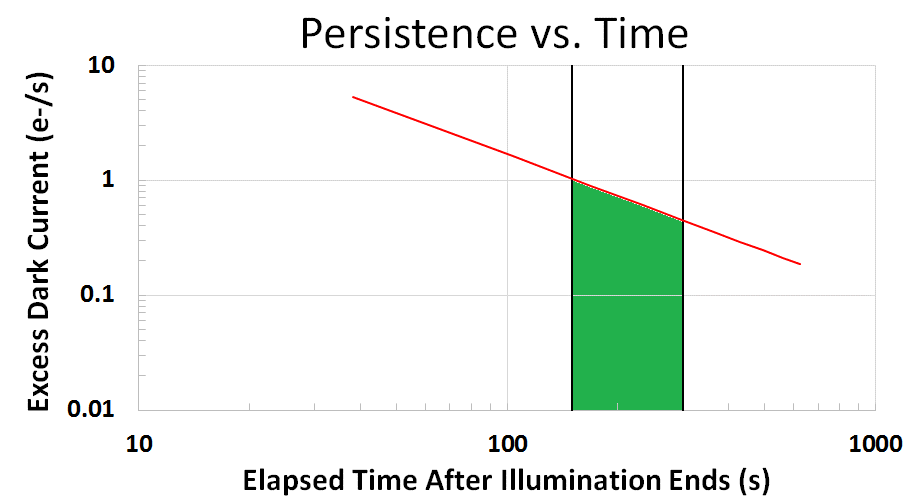} 
\end{subfigure}
\caption{Illustration of persistence calculation. In the sample plot above (log-log scale), the integrated signal (green area) under a power law fit (red line) results in accumulation of 100 e- between 150 and 300 seconds after detector reset.}\label{fig:per-meas}
\end{centering}
\end{figure*}

The dark exposures following the illuminated exposure are dark current corrected, and the amount of persistence is calculated by fitting a power law to the excess dark current present in the 150~s that occur after a 150~s interval following illumination. This power law fit is then integrated to calculated the integrated persistence in a 150~s exposure. Figure~\ref{fig:per-meas} illustrates this persistence calculation scheme.

\begin{figure*}[h]
\begin{centering}
\begin{subfigure}[t]{0.45\textwidth}
        \centering
        \includegraphics[width=0.7\textwidth]{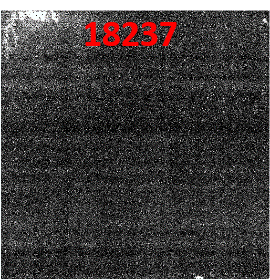} 
        \caption{Persistence image for SCA 18237.} \label{fig:sca-18237-per-img}
\end{subfigure}
\begin{subfigure}[t]{0.45\textwidth}
        \centering
        \includegraphics[width=0.7\textwidth]{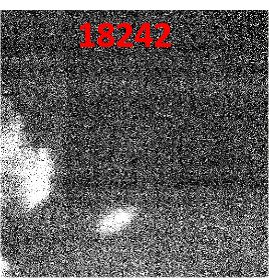} 
        \caption{Persistence image for SCA 18242.} \label{fig:sca-18242-per-img}
    \end{subfigure}
\caption{(a) An example of the persistence image for the H4RG-10 device 18237. (b) An example of the persistence image for the H4RG-10 device 18242.  Using these images, we can estimate the median and per pixel persistence performance of the H4RG-10 devices. SCA 18237 has a more uniform level of persistence. SCA 18242 has more structure with two large areas of enhanced persistence, but both arrays have acceptable median persistence values of 0.02~\%FW. }\label{fig:per-img}
\end{centering}
\end{figure*}

The goals for the detector program were to produce detectors with less than about 0.1~\%FW persistence in a 150~s exposure that occurs 150~s after the detector is reset following illumination, where FW is $\sim$100~ke. All but one of the full array devices tested with the PV3 passivation satisfy this benchmark. The median value of persistence 150~s after intense (800~ke-) illumination is 0.02~\%FW. This performance exceeds the goal, and for comparison, represents a factor of 20 improvement in performance compared to the Hubble Space Telescope's Wide Field Camera 3 detectors ($\sim$0.4\%FW in a 150~s exposure). Unsurprisingly, there is still spatial variation of the persistence across individual detectors (see Figure~\ref{fig:per-img}). We have observed that this spatial variation of the persistence does depend on the level of illumination as well. Detectors that show high spatial variation after intense illumination have shown more uniform persistence for sub-full well illumination. See Table~\ref{tab:summary}, for a summary of performance characteristics.

\subsection{Quantum efficiency performance}
Quantum efficiency (QE) is a measure of the fraction of incident photons on a detector that get detected as signal. High QE is therefore essential for a sensitive and efficient detector system. The QE will depend on the specific detector material, coatings, and pixel design and consequently will vary depending on the wavelength of light.

\begin{figure*}[h]
\begin{centering}
\begin{subfigure}[t]{\textwidth}
        \centering
        \includegraphics[width=\textwidth]{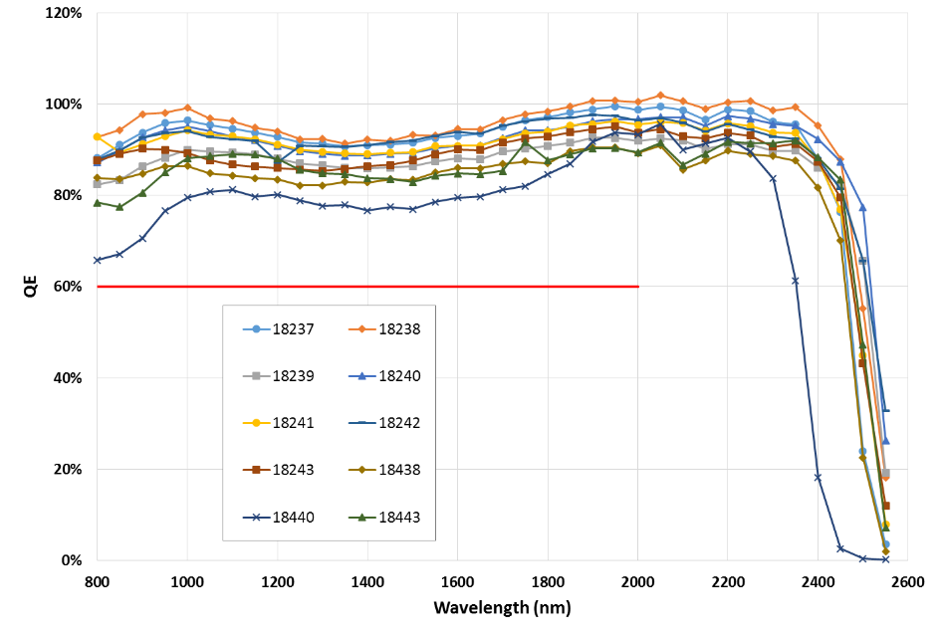} 
\end{subfigure}
\caption{Plot of the measured quantum efficiency of the detectors in the PV3 full array lot of WFIRST's H4RG-10s. The QEs between 0.8 -- 2.3$\mu$m all exceed 60\%. The estimated error in monochromatic results is $\sim$5\%.}\label{fig:qe-img}
\end{centering}
\end{figure*}

The quantum efficiency is measured by comparing the measured signal of a detector under test to the estimated signal incident on the detector. The signal incident on the detector is estimated by using a transfer curve between a calibrated photodiode from the National Institute of Standards and Technology (NIST) at the detector position and a secondary photodiode in the test dewar. To measure the quantum efficiency of a detector, we illuminate the detector with a uniform and monochromatic (2--50~nm bandpass) light source. We acquire at least 2 exposures with 5 samples up-the-ramp for each wavelength, targeting a total signal of $\sim$10000~e. The signal level from the secondary photodiode is also recorded during these exposures. The exposures are averaged to form single wavelength images of measured signal. The measured signal images are then divided by the estimated incident signal that is calculated using the secondary photodiode transfer curve and the exposures' measured signal from the secondary photodiode. The ratio of the measured signal and the estimated incident signal forms a QE image for each wavelength. The QE curve is presented as the global median across these QE images as a function of wavelength in Figure~\ref{fig:qe-img}.

The average quantum efficiency from 0.8 -- 2.3~$\mu$m for all the full array PV3 devices is above 60\%, with a median average of 91\%. See Table~\ref{tab:summary}, for a summary of performance characteristics. The typical uncertainty in the monochromatic QE measurements is estimated at $\sim$5\% with the precision of the QE measurements (repeatability) accounting for 2\%. The QE measurement methodology continues to be reviewed in collaboration with NIST to minimize the measurement uncertainties.

\subsection{Linearity performance}
The raw signals from the WFIRST H4RG-10s in a detector system, as with many IR photodiode detectors, will be subject to nonlinearities (e.g., classical nonlinearity, burn-in, reciprocity failure) as highlighted in \S\ref{sec:theory:traps}. To successfully use data from these detectors, the signal must be linearized via a linearity calibration. In a typical calibration, a pixel's signal up-the-ramp is fit with a high order polynomial to characterize its nonlinearity. This high order fit is then used to correct for a pixel's CNL (see \eg Refs.~\citenum{Hilbert-2014,Canipe-2016}). Stable pixels with fewer defects are expected to behave more regularly (and thus more linearly), contributing less to measurement error and systematics. In contrast, pixels with more numerous defects may be less stable due to trapping mechanisms described in \S\ref{sec:theory:traps}. Therefore, it is important to quantify the linearity and the precision to which pixels can be linearized for our arrays.

Linearity is measured for each pixel in the array under test and quantified by the maximum percentage deviation of a pixel's signal from a high order polynomial fit to the signal. In this sense, the linearity measurement here is more aptly described as a measurement of the ability of a pixel to be linearized using this technique.  The degree to which a pixel's signal can be fit using a polynomial of fixed order is then a measure of the effectiveness of current techniques to correct that pixel's ramps. Linearity measurements are taken using flat-field illumination with a 1.4~$\mu$m source. We acquire 50 exposures with 50 samples up-the-ramp of the illuminated detector. The exposures are averaged, and then the signal up-the-ramp per pixel is fit using a 6th degree polynomial. The linearity measurements are calculated in the region of the ramp between 0 and 80~ke- worth of collected signal. Above 80~ke-, most of the photodiodes approach full well. The polynomial fit of a pixel's ramp is subtracted from the pixel ramp to generate a residual vector. This residual vector is divided by the polynomial fit and multiplied by 100 to generate a percentage deviation vector. The maximum of this vector in the signal regime of $\sim$0 -- 80~ke- is the maximum percentage deviation. We refer to these values as the maximum normalized residuals (MNR). 

\begin{figure*}[h]
\begin{centering}
\begin{subfigure}[t]{0.45\textwidth}
        \centering
        \includegraphics[width=0.8\textwidth]{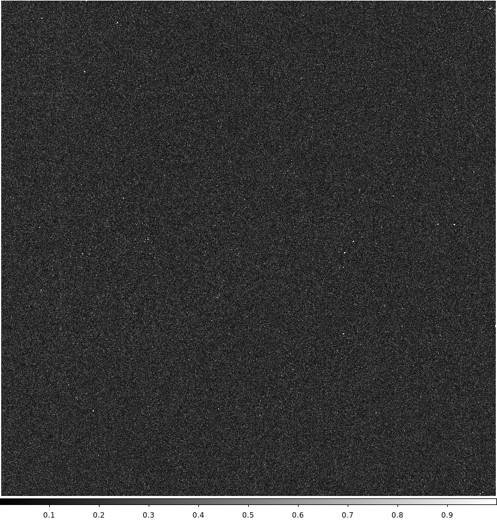} 
        \caption{MNR image for 20829.}
\end{subfigure}
\begin{subfigure}[t]{0.5\textwidth}
        \centering
        \includegraphics[width=\textwidth]{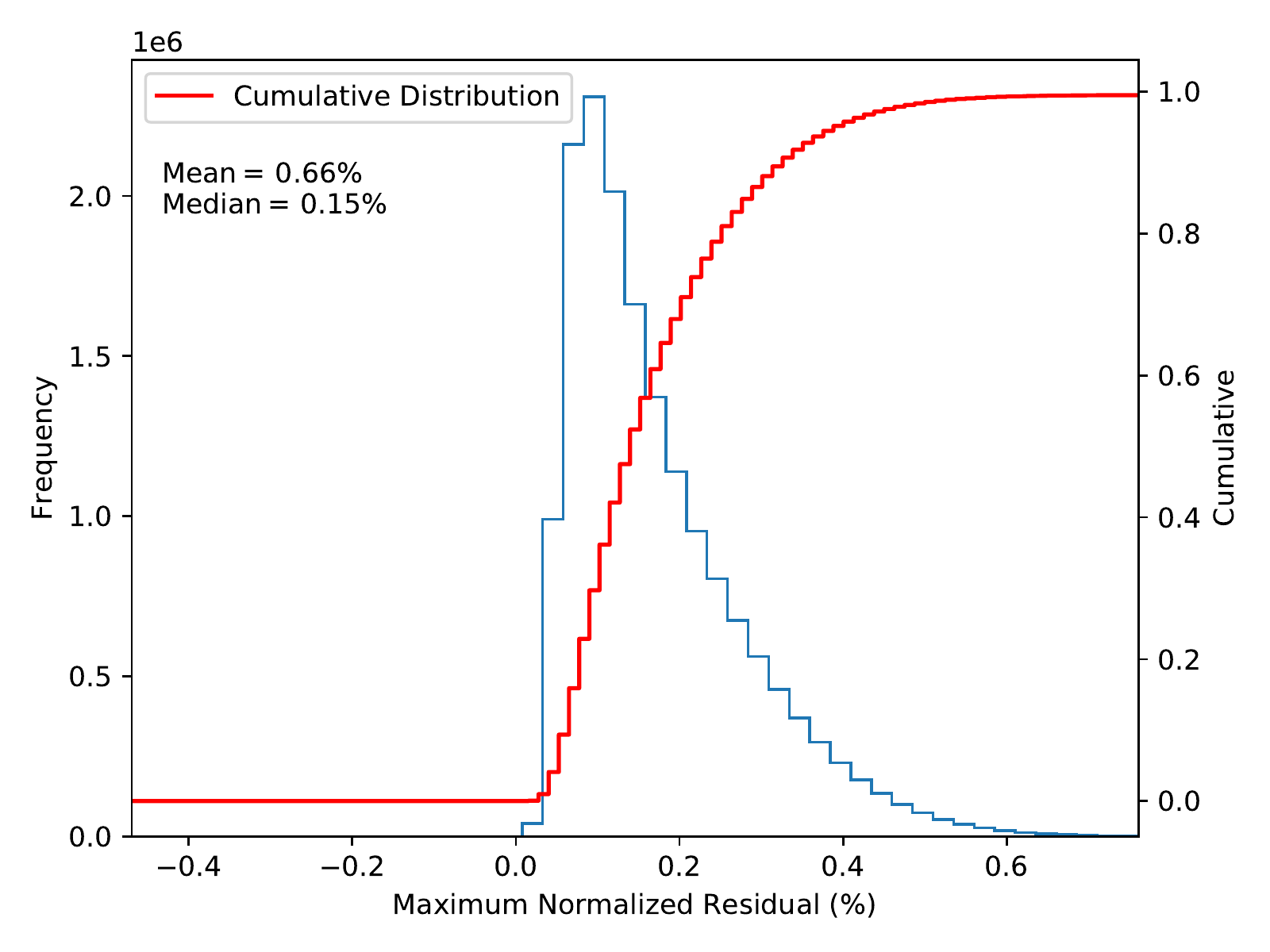} 
        \caption{MNR histogram for 20829.} \label{fig:lin-meas-hist}
    \end{subfigure}
\caption{Image of the maximum normalized residual (MNR) for flight candidate SCA 20829. SCA 20829 used the same passivisation as the PV3 full array lot. Nearly each active pixel (99.9\%) have residuals to the polynomial fit up to 80~ke- less than 1\%.}\label{fig:lin-meas}
\end{centering}
\end{figure*}

Linearity was not measured for the full array PV3 lot devices, so we present results from a PV3 flight candidate. Figure~\ref{fig:lin-meas} shows the 2-D image and histogram of these maximum normalized residuals for the flight candidate array SCA 20829. Nearly every pixel has a MNR value less than 1\%. Though these measurements are taken below full well at 80~ke-, measurements from the detector above 80~ke- can be useful. These measurements, however, will likely have higher linearity corrections.

\subsection{Interpixel capacitance and cross talk performance}\label{sec:crosstalk}

The origin of interpixel capacitance (IPC) is at the point of charge collection Ref.~\citenum{Moore-2006}. Crosstalk more generally can also occur via diffusion before charge carriers are generated and in the readout electronics. See theory Section~\ref{sec:theory:spreading} above where charge spreading is more thoroughly discussed. But we combine these two properties here since in practice we measure the sum effect of both. This cross-talk can degrade measurements of source shapes (e.g. for PSF measurements\cite{Kannawadi-2016}), so ideally crosstalk should be minimized and precisely characterized.

\begin{figure*}[h]
\begin{centering}
\begin{subfigure}[t]{0.45\textwidth}
        \centering
        \includegraphics[width=0.7\textwidth]{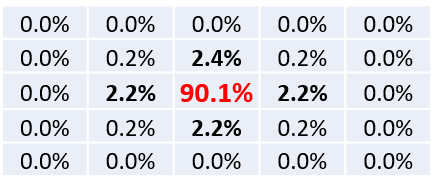} 
        \caption{IPC and crosstalk matrix for SCA 18237.} \label{fig:sca-18237-ipc}
\end{subfigure}
\begin{subfigure}[t]{0.45\textwidth}
        \centering
        \includegraphics[width=0.7\textwidth]{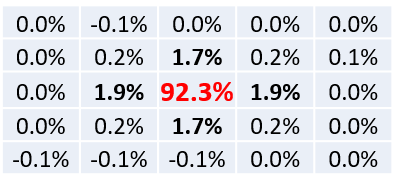} 
        \caption{IPC and crosstalk matrix for SCA 18242.} \label{fig:sca-18242-ipc}
    \end{subfigure}
\caption{(a) An example of the IPC and crosstalk matrix for the H4RG-10 device 18237. (b) An example of the IPC and crosstalk matrix for the H4RG-10 device 18242. The mean percent of signal coupled to the nearest neighboring pixels for both arrays is $\leq$3\% and within the range of acceptable performance. These measurements were made using a $^{55}$Fe source and thus include effects from both the readout electronics and charge diffusion in the detector material.}\label{fig:ipc-mat}
\end{centering}
\end{figure*}

We measure the interpixel capacitance and crosstalk in two ways. In both methods, the normalized signals in the four nearest neighbors of a ``hot" pixel are calculated and used to infer the percent of the signal incident on a single pixel that is coupled to neighboring pixels. The first method uses the hits from a radioactive $^{55}$Fe source\cite{Fox-2009}. With an $^{55}$Fe source in the dewar, we acquire at least 3 exposures analogous to the dark current data set (i.e. total exposure time 7258~s, 101 samples up-the-ramp). The detector absent of any illumination detects X-ray events, identified by their symmetry and amplitude above the background. Postage stamps of 7x7 pixels centered on these hits are identified in the exposure frames. The mean of the outer ring of pixels is subtracted from the inner 5x5 pixel grid, and then all the 5x5 pixel grids are averaged together to form an average 5x5 pixel grid. The pixel values in the 5x5 grid are summed and the result is used to normalize each pixel value. The normalized 5x5 grid is then multiplied by 100 to generate a 5x5 grid of the crosstalk by percent. Figure~\ref{fig:ipc-mat} shows the 5x5 pixel grids generated for two typical arrays to estimate the percent of the coupled signal.

The second method uses single pixel reset. To measure the IPC from the readout electronics (excluding charge diffusion in the detector material) we acquire two exposures. The unilluminated detector is held in reset for the first exposure. A second exposure is then acquired with every 8th pixel (in column and row) reset to a reset voltage typically 0.3V above the normal reset voltage. These two exposures are subtracted, and the 7x7 pixel grids centered on the reset pixels analyzed in the same way above. The background from the outer ring is subtracted from the inner ring. The 5x5 grids are then averaged together and normalized by their sums and multiplied by 100 to generate 5x5 grids of the percentage of signal from a pixel to its nearest neighbors. The IPC without charge diffusion for the full array lot PV3 detectors was not measured. But the single pixel reset results are being measured for each detector in the flight lot.

A benchmark of the detector development program was to achieve less than 3\% on average of a pixel's signal coupled to its nearest neighbors. Every detector tested with the PV3 passivation achieves this goal. The median value of the sum effect of interpixel capacitance and crosstalk across the devices tested is 1.9\%. See Table~\ref{tab:summary}, for a summary of performance characteristics.

\begin{table}
\centering
\resizebox{\linewidth}{!}{%
\begin{threeparttable}
\caption{Summary of detector performance for full array lot with PV3 passivation.}
\begin{tabular}{lcccccc}
\toprule
Detector & Median CDS Noise (e-) & Total Noise (e-) & Median Dark Current (e/s) & Persistence (\% ) \tnote{a}&   QE (\%) \tnote{b}& Crosstalk (\%)  \\
\midrule
\rowcolor{Gray}
Benchmark      &                  $<$ 20 &     N/A          &                     $<$ 0.1 &           $<$ 0.10 &     $>$ 60 &            $<$ 3 \\
   18237 &                  11.9 &             5.56 &                     0.001 &             0.02 &       95 &            2.3 \\
   18238 &                  15.1 &             7.03 &                     0.001 &             0.01 &       96 &            2.6 \\
   18239 &                  15.2 &             6.72 &                     0.001 &             0.03 &       89 &            1.8 \\
   18240 &                  15.7 &             6.99 &                     0.001 &             0.01 &       93 &            2.3 \\
   18241 &                  15.2 &             7.12 &                     0.004 &             0.02 &       92 &            1.9 \\
   18242 &                  16.0 &             8.01 &                     0.040 &             0.02 &       93 &            1.8 \\
   18243 &                  16.3 &             8.65 &                     0.064 &             0.03 &       90 &            1.8 \\
   18244 &                  15.1 &             7.03 &                     0.003 &             0.20 &       90 &            1.9 \\
   18438 &                  13.2 &             7.13 &                     0.001 &             0.01 &       86 &            1.9 \\
   18440 &                  14.4 &             7.20 &                     0.001 &             0.01 &       81 &            2.4 \\
  18441* &                   --- &              --- &                       --- &              --- &      --- &            --- \\
   18442 &                  16.2 &            11.00 &                     0.350 &             0.02 &       93 &            1.9 \\
   18443 &                  12.8 &             6.38 &                     0.003 &             0.02 &       87 &            2.2 \\
    Mean &               14.8 &          7.40 &                 0.039 &        0.03 &  90 &    2.1 \\
  Median &                 15.2 &            7.08 &                     0.002 &             0.02 &       91 &            1.9 \\
   Sigma &               1.4 &          1.36 &                 0.100 &        0.05 &  4.2 &    0.3 \\
\bottomrule
\end{tabular}

\begin{tablenotes}
\small
\item Notes: The gray highlighted row lists the respective goals of the detector full array production runs. For example, the benchmark for median CDS noise was $<$20~e-.
\item[a] Persistence in \% of full well in a 150 second exposure, after a 150 second interval after illumination ends.
\item[b] Quantum efficiency averaged over 0.8--2.3$\mu$m.
\item[*]SCA 18441 was discovered to have a coupling between two detector bias voltages and was not tested further.

\end{tablenotes}
\label{tab:summary}
\end{threeparttable}
}
\end{table}

\subsection{Brighter-fatter effect performance}
BFE describes the observed dependence of pixel sensitivity to accumulated charge as described in \S\ref{sec:theory:BFE}. This signal dependent sensitivity affects the systematics and precision of shape measurements needed for the HLIS science cases. The dependence of the system PSF with signal level will also generally be important for investigations using PSF-fitting photometry. Thus, the magnitude of this effect must be characterized. 

The measurement of BFE in the full array lot data was not an explicit benchmark or goal of the detector development program, but the DCL in collaboration with members of the WFIRST science team have been able to begin characterizing this property of WFIRST's infrared arrays. As detailed in Refs.~\citenum{Hirata2020,Choi2020}, an estimate of the BFE can be made using sets of flat fields and careful calculation of a non-overlapping correlation function of this data in ``super pixels" across the array to achieve better signal-to-noise. SCA 18237 was analyzed using this technique in Ref.~\citenum{Choi2020} using \textsc{solid-waffle}, and the authors found a coupling coefficient to the four nearest neighbors of 0.287 $\pm$ 0.003 ppm/e. This coupling coefficient describes how charge from neighboring pixels impact the effective pixel area of a given pixel. In light of the much smaller estimates of the nonlinear IPC (i.e. the dependence of IPC on signal level) estimated for this array, the BFE is likely responsible for the bulk of this coupling. In the current absence of additional estimates of BFE in the full array lots, this measurement provides a baseline estimate for the full array performance. 

\subsection{Performance after thermal cycling, vibration, and radiation}

The WFIRST H4RG-10 design has been successfully qualified following thermal cycling, vibration, and radiation testing\footnote{See \url{https://wfirst.gsfc.nasa.gov/science/sdt_public/wps/references/WFIRST_DTAC5_nobackup.pdf} for additional details}. Thermal and vibration testing took place in the DCL. SCA 18237 was subjected to 40 thermal cycles between 295~K and 100~K and subsequently underwent vibration testing at General Environmental Verification Specification (GEVS) workmanship levels. No degradation in performance at 100~K and 1~V bias operation was observed.

Radiation testing took place at the UC Davis cyclotron. SCA 18239 was exposed to an equivalent of 5~krad of total ionizing dose (fluence of $3\times10^{10}$/cm$^{2}$ of 63~MeV protons). Measurements of dark current and noise were taken immediately following irradiation and after the array was returned to Goddard. Immediately after irradiation, the DCL tested the performance of SCA 18239 and found additional hot pixels (3.5\%) and an increase in the median dark current (0.003~e-/s to 0.012~e-/s). The additional hot pixels are within the desired operablility specification and expected. The increase in median dark current is still modest enough to be negligible in a typical WFIRST 150~s exposure. In testing at the DCL after irradiation, noise performance was essentially unchanged except for the addition of $\sim$7.3\% pixels with noise above 8~e-.

\subsection{Flight lot detector performance}
The production of the flight lot WFIRST H4RG-10 detectors started in 2019 using the PV3 passivation recipe developed in the WFIRST detector development program. After a cold functional screen test at Teledyne, detectors are hand carried to the Goddard DCL. Flight lot detectors undergo acceptance testing in the DCL to help sort and rank performance of individual arrays. The flight lot detectors that pass acceptance testing are deemed flight candidate detectors. Flight candidates are then further characterized to ensure sufficient data for the detectors exist to supplement flight calibration activities for the WFIRST mission.

\begin{figure*}[h]
\begin{centering}
\begin{subfigure}[t]{\textwidth}
        \centering
        \includegraphics[width=\textwidth]{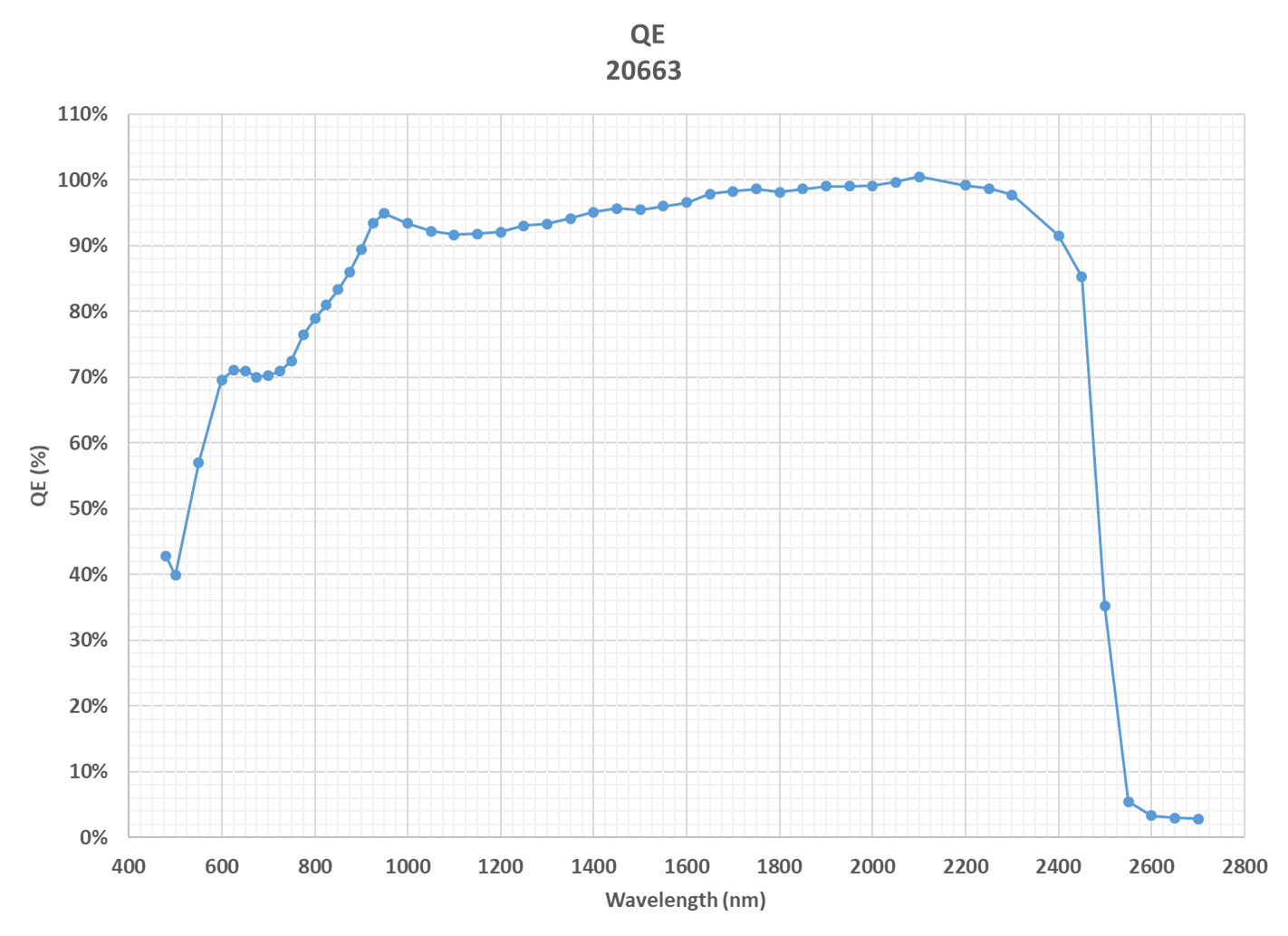} 
\end{subfigure}
\caption{Plot of the measured quantum efficiency of a flight candidate array, extending measurements to bluer wavelengths. The estimated error in monochromatic results is $\sim$5\%.}\label{fig:flight-qe}
\end{centering}
\end{figure*}

The flight acceptance tests cover much of the detector performance checks described in \S\ref{det-per}. However, for flight detectors, the performance requirements in general are more thoroughly specified. For example, persistence requirements are specified for two initial stimulus values (50~ke- and 300~ke-) versus the single 800~ke- stimulus used in the performance tests described for the full array lots. Flight acceptance tests also include extended quantum efficiency measurements that extend to bluer wavelengths (see Figure~\ref{fig:flight-qe}). In addition, the acceptance tests more closely mirror the operation of the WFIRST detectors in flight, operating at temperature of 95~K. As of April 2020, thirty-two devices have passed through acceptance testing in the DCL, twenty-four of these devices pass the Teledyne cold functional screen test. Six of these devices have passed the flight acceptance tests to become flight candidates with a yield of 25\% of flight candidates at acceptance testing. This yield is on target to provide the required 18 detectors out of the expected 72 delivered passing Teledyne's screening test by August 2021.

The results from acceptance tests of the current set of flight lot detectors are similar to the full array lot. Noise performance, dark current performance, and persistence performance continue to be in line with the goals and milestones of the detector development program. The first few flight detectors are in the process of beginning characterization tests as well. A preliminary summary of the noise, dark current, and persistence performance of five flight candidate detectors is shown in Table~\ref{tab:curr-flight}.

\begin{table}
\centering
\begin{threeparttable}
\caption{Summary of detector performance for 5 WFIRST flight candidates with PV3 passivation.}
\begin{tabular}{lcccc}
\toprule
{} & Total Noise & Dark Current & \multicolumn{2}{c}{Persistence (300 ke)\tnote{a}} \\
{SCA} &  Median (e) & Median (e/s) &                   Median (e/s) & (\%) Pixels passing \\
\midrule
20849       &        5.41 &        0.001 &                          0.173 &              99.90 \\
20829       &        5.73 &        0.000 &                          0.207 &              99.96 \\
20828       &        5.34 &        0.000 &                          0.144 &              99.90 \\
20663       &        5.48 &        0.000 &                          0.100 &              99.88 \\
20833       &        4.97 &        0.000 &                          0.074 &              99.93 \\
\midrule
Requirements&       $<$6.5 &    $<$0.05    &    $<$0.50    & $>$95\%\\    
\bottomrule
\end{tabular}
\begin{tablenotes}
\small
\item Notes: Total noise and median dark current are calculated as described in \S\ref{det-per}.
\item[a] Median persistence in a 150 second exposure, 10 minutes after illumination. Percent pixel passing is percentage of pixels with persistence $<$0.8 e-/s.
\end{tablenotes}
\label{tab:curr-flight}
\end{threeparttable}
\end{table}

\section{Summary}

The WFIRST mission has taken on the ambitious task to execute a broad scientific program to address key questions in astronomy and astrophysics. These include what is the nature of the cosmic expansion of the universe and how does it evolve and how do planets form and what are the demographics of extrasolar planets? WFIRST aims to answer these questions using four unique surveys in addition to supporting guest observations and an archive program. At the heart of the WFIRST observatory are 18 HgCdTe detectors of the Wide Field Instrument. To meet the challenging tasks of minimizing measurement error and producing rich data sets free from large systematics, the WFIRST mission undertook a detector development program to produce WFIRST's H4RG-10 detectors. The detector development program was phased and involved the manufacturing of banded arrays to select a pixel architecture and design to optimize key performance metrics such as noise and persistence. 

We have provided an overview of the detection process for the WFIRST H4RG-10 including, for example, theoretical estimates of the absorption depth of photons in the array and the extent of charge diffusion. We have also presented the results of the full array lot of the selected pixel architecture and design that uses a PV3 passivation. The full array lot detectors as an ensemble meet the benchmark goals of the detector development program for low noise, low persistence, and good quantum efficiency. Testing and characterization of the flight lot detectors has begun, and we have presented some preliminary results for the first few flight candidate detectors. Like the full array lot, the first flight lot candidates exhibit low noise, low dark current, and good persistence performance.

\section{Acknowledgments} 

This work was supported by NASA as part of the Wide Field Infrared Survey Telescope (\WFIRST) Project. We wish to thank our \WFIRST colleagues, many of whom carefully read the paper and provided invaluable feedback. These include Sebastiano Calchi Novati of the Infrared Processing and Analysis Center (IPAC), Chris Hirata and Ami Choi of Ohio State University, John MacKenty of the Space Telescope Science Institute (STScI), Chaz Shapiro of NASA JPL, and Mario Cabrera of Alcyon Technical Services LLC.

{\it Software: Astropy,\cite{Astropy1,Astropy2} Mathematica,\cite{Mathematica} Matplotlib,\cite{Matplotlib} NumPy,\cite{Numpy} Pandas,\cite{Pandas} SciPy.\cite{SciPy}}


\bibliography{refbib}   
\bibliographystyle{spiejour}   

%
\vspace{2ex}\noindent\textbf{Gregory Mosby, Jr.} is a Research Astrophysicist and detector scientist at NASA Goddard Space Flight Center. He received his BS degree in astronomy and physics from Yale University in 2009, and his MS and PhD degrees in astronomy from the University of Wisconsin - Madison in 2011 and 2016, respectively. His current research interests include near infrared detectors, astronomical instrumentation, and applications of machine learning to observational astronomy. He is a member of SPIE.

\vspace{2ex}\noindent\textbf{Bernard J. Rauscher} is an experimental astrophysicist at NASA's Goddard Space Flight Center. His research interests include astronomy instrumentation, extragalactic astronomy and cosmology, and most recently the search for life on other worlds. Rauscher's work developing detector systems for \JWST was recognized by a (shared) Congressional Space Act award and NASA's Exceptional Achievement Medal.

\vspace{1ex}
\noindent Biographies of the other authors are not available at this time.

\listoffigures
\listoftables

\end{spacing}
\end{document}